\documentstyle[aps,preprint,psfig]{revtex}
\begin{document}
\title{\bf Traffic Equations and Granular Convection}
\author{ Daniel C. Hong \thanks{
E-mail address: dh09@lehigh.edu}
and Su Yue \thanks{E-mail address: sufrank.yue@citicorp.com}}            
\address{
Department of Physics, Lewis Laboratory,
Lehigh University, Bethlehem, Pennsylvania 18015}
\date{today}
\maketitle
\begin{abstract}
We investigate both numerically and analytically the convective
instability of granular materials by two dimensional traffic equations.
In the absence of vibrations traffic
equations assume two distinctive classes of fixed bed
solutions with
either a spatially uniform or nonuniform density profile.  The former one
exists only when the function $V(\rho)$ that monitors the relaxation of
grains assumes a cut off at the closed packed density, $\rho_c$, with
$V(\rho_c)=0$, while the latter one exists for any form of V.  
Since there is little difference between the uniform and nonuniform
solution deep inside the bed,
the convective instability of the bulk may be studied by
focusing on the stability of the uniform solution.  In the presence of
vibrations, we find that the uniform solution bifurcates into a
bouncing solution, which then undergoes a supercritical bifurcation
to the convective instability.  We determine the onset of
convection as a function of control parameters and confirm this picture by
solving the traffic equations numerically, which reveals bouncing solutions,
two convective rolls, and four convective rolls.  Further, convective
patterns change as the aspect ratio changes: in a vertically long
container, the rolls move toward the surface, and in a horizontally long
container, the rolls move toward the side walls.  
We compare these results
with the those reported previously with a different continuum model by
Hayakawa,Yue and Hong[Phys. Rev. Lett.{\bf 75}, 2328 (1995)].
Finally, we also present a derivation of the traffic equations from Enskoq
equation.                                                            
\vskip 0.2 true cm
\noindent P.A.C.S. numbers: 47.20.-k,46.10.+z,81.35.+k

\end{abstract}
\vskip 1.0 true cm
\noindent {\bf I. Introduction}
\vskip 0.2 true cm
This paper is concerned with the numerical as well as the analytical 
analysis of the 
two dimensional traffic equations, termed model B in the literature [1] with
an aim to understand the convective instability of granular materials. 
It has been known since Faraday [2] that the
granular materials in a vibrating bed develop permanent convective rolls
when the strength of the vibration exceeds a critical value.  Unlike
the well known Rayleigh-Bernard thermal convection in fluids, however, the 
origin of the granular convection has remained relatively unexplored
since its discovery, but recently
two simultaneous push from experimental side [3,4]
with the use of MRI
or X-ray method as well as  from computer simulations
based on the distinct
element method [5] have substantially aided our understanding
through visualization.  Yet, the theoretical efforts to uncover the basic 
mechanism of this granular convection have not been remarkable, still largely
focused on producing convective patterns through computer models and
simulations.  While many impressive experimental data are currently being 
piled up, theoretical development [6-9] 
in this area seems to be still far from complete.
In order to have a deeper understanding of the granular convection,
we find it essential to come up with a reasonable continuum model that
contains a minimum number of control parameters yet
captures some of the essence of granular convection.  
The goal is certainly to explore analytically and
numerically the instability mechanisms that lead to many fascinating complex
nonlinear dynamics behaviours.

Our aim here is two fold: we first present such a continuum model
along with previously unpublished numerical
results based on two dimensional traffic equations,
and second, we carry out the linear stability analysis of the
traffic equations and uncover the mechanism of granular convection as a 
supercritical bifurcation of a bouncing solution. 
This mechanism for the granular convection is essentially identical to
the previously obtained scenario for a
different continuum model, termed model A [8].  However, detailed simulations
of these two continuum models reveal distinctively different convection
patterns for different geometries and parameter ranges.  Most
notable departure between the two is the migration of
convective rolls toward the surface and toward the walls in model
B, as the aspect ratio changes.  This is in sharp contrast with
the results of the model A that predicts a series of
rolls in the container.  Further, the simulation results of
the model B
yield rich dynamical behaviors as we probe deeper into the parameter space,
yet the analytical structure of the nontrivial patterns 
seem to be quite complex and require
further extensive studies in the future.
Our analytical study here focuses exclusively on the stability analysis
of a uniform bouncing solution, but there exists a second class of
solutions that are spatially nonuniform.  The stability analysis of the latter
is highly nontrivial and we only present a brief description of it in the
Appendix B.  Fortunately, however, since there is
little difference between the uniform and nonuniform solutions deep inside
the bed, we expect that the uniform bouncing solution may capture the
essence of the bulk instability.  For details, see the text.

This paper is organized as follows. We first define the model equations in II
and make attempt to derive the traffic equations from the Enskoq equation
in the Appendix A.
We will then present numerical results in II, and provide some insight
into the stability mechanism of the granular convection in III.  We also
discuss several unresolved questions in IV.  We now turn to the main text

\vskip 0.5 true cm
\noindent {\bf II. Two Dimensional Traffic Equations and Granular Convection}
\vskip 0.2 true cm
Apart from its wide application
to traffic flow problems [12], the traffic equations [13-15]
have recently been proposed, as an alternative continuum model 
to Navier-Stokes equation, in investigating a variety of 
dynamical responses
of granular materials.  Examples include the
granular relaxation under repeated tapping [16], density waves and
jamming and clogging phenomena [11,14,17], 1/f spectrum of hopper flow [18],
and granular segregation [19].
But we point out that, unlike the studies to be presented in this paper,
almost all the studies thus far have focused 
exclusively on
{\it one} dimensional traffic equations [13-19], 
which certainly will not be able to
describe the granular convection.  Hence, we first present here the 
two dimensional version of the traffic equations [1], termed model B in 
ref.[1]:
\begin{eqnarray}
\label{traffic_c}
 {\partial{\rho} \over \partial{t}} + {\partial{\rho v_x} \over \partial{x}}
+ {\partial{\rho v_z} \over \partial{z}} = 0 \label{mass} \\
 {\partial{v_x} \over \partial{t}} + v_x {\partial{v_x} \over \partial{x}}
+ v_z {\partial{v_x} \over \partial{z}}  
=   - c_0^2/\rho  {\partial{\rho} \over \partial{x}} 
+ \mu ({\partial^2 v_x \over \partial x^2} 
 + {\partial^2 v_x \over \partial z^2}) \\
{\partial{v_z} \over \partial{t}} + v_z {\partial{v_z} \over \partial{z}}
+ v_x {\partial{v_z} \over \partial{x}}  
=   {V(\rho,t) - v_z \over \tau} - c_0^2/\rho  {\partial{\rho} \over 
\partial{z}}
+ \mu ({\partial^2 v_z \over \partial x^2} 
 + {\partial^2 v_z \over \partial z^2})
\end{eqnarray}

\noindent where $c_o^2\sim T_e$ is the sound speed, and $\mu$ is the
viscosity.  The difference between
the model B [1] and the model A [8] is the presence of the relaxation
term in the z direction, which is represented by an average function
$V(\rho,t)$ with the relaxation time $\tau$. 
The net effect is for the void(or particle) to adjust its speed, $v_z$,
around the average value $V(\rho)$ in a given time $\tau$. 
The origin of
such a term has been discussed in Hong et al [16] in an
attempt to introduce correlations among grains or voids in the diffusing
void model(DVM)[20].  In the DVM, the void speed is only a function
of local void density, namely $v_z=V(\rho)$ + diffusion term.  However, a
void is a compressible hydrodynamic object that changes and adjusts its shape
to conform to the surrounding, not instantaneously, but in a given
finite time, $\tau$.  So, it may be appropriate to write down the time
dependent equation for the velocity in a manner given by (3) than simply 
assuming a fixed value at a given local density.  The presence of such
relaxation process may be effectively equivalent to assuming a drag force
acting on a void.

The coupled equations (1) -(3) are fairly similar to the two-phase model [26]
of a fluidized bed that has been widely used for mixtures
of gas and granular particles.  As shown in 
the Appendix A, functions in the traffic equations (1)-(3)
may be inferred from the Enskoq equation [21]; 
namely $-v_z/\tau$ is the drag term 
imposed externally on the particle, or by vibrations.  In the case of
no interstitial fluid, its origin lies in the frictions of the front and the 
rear glass of the container and from the wall.
Further, the Enskoq pressure, 
$T\rho(1+f(\rho)\rho/2)$ with $f(\rho)$ the correlation function, produces an 
extra term $V(\rho)$ in addition to the hard sphere pressure $T\rho$.  In this
case, we make an important observation that
the coefficient of $V(\rho)$ is proportional to the gravity g, which will
enable us to incorporate the vibrations of the bed.
This observation makes sense, because the strength of the 
mean speed(also termed the drift velocity) is determined by the gravity and
thus it is quite physical to assume that the mean speed of the void is also
proportional to the gravity as demonstrated in [16] and Appendix A.
In the moving frame of reference of the vibrating bed, the mean speed
$V(\rho)$ depends not only on the density but also on the way the box is
shaken.  In the moving frame, the effective gravity,
 $g^{'} = - g + A \omega^2 cos(\omega t)$,
is the only time dependent factor involved in the mean speed $V(\rho)$.
Therefore, without the loss of generality, we expect that
the mean speed of the granular flow along the $z$ direction 
assumes the following form:

\begin{equation}
V(\rho) = V_0(\rho)\cdot g^{'} = V^{'}_0(\rho) \cdot (-1 + \Gamma sin( \omega t))
\end{equation}
\noindent 
where $\Gamma = A \omega^2 / g$ with $A$ the amplitude of the 
vibration, and $\omega$ the vibrational frequency.
$V_0(\rho)$ is the mean speed in the absence of the vibration, such
as the case in the hopper flow.
In one dimensional traffic equation, we have used [16]:

\begin{equation}
V_0(\rho) = \alpha / (1 + \beta \rho^{2})
\end{equation}

\noindent
Note that the functional form 
of $V_o$ is not unique, and other exponentially decaying functions
have also been proposed [13].
The $V_0(\rho)$ assumed above has a long tail when the density increases,
which might be unphysical since there should be a critical
density, $\rho_c$,
for any granular material, at which  all particles will be locked in if
the density exceeds
that critical density.  This critical density may be proportional
or equal to the closed packed density.  Thus, we have employed a different
form in our numerical simulations that assume the form (Fig.1a):

\begin{equation}
\label{drag}
V(\rho)
=(\rho_c-\rho)^{\beta}\theta(\rho_c-\rho)
\end{equation}
where we have used a
step function $\theta(\rho_c-\rho) =0$ when $\rho \geq \rho_c$.
A physically reasonable
choice of $\beta$ may be such that $V'(\rho)<\infty$.  (For
$\beta <1$, $V'(\rho)$ diverges at $\rho_c$, and thus in our stability
criterion, we set the density of the uniform solution slightly less than
the closed packed density.)
Before investigating eqs.(1)-(3) numerically, we first check the consistency.

{\bf A. Consistency check: static solutions without vibrations}.
It is worthwhile to examine a few special cases not only for the sake of
general limitation, but also to check the numerical accuracy.
When $\Gamma = A \omega^2/g = 0$, which corresponds to the case without 
the vibration, which we also term ``the fixed bed solution'', we
expect a static solution where $\vec{v} = 0$ in all momentum equations.
We note that there are {\it two} classes of static solutions.

First, a spatially nonuniform solution.  Regardless of the functional form
of $V(\rho)$, we find by inspection that  
there exists a spatially nonuniform static solution that satisfies
the following differential equation:

\begin{equation}
\label{static}
V_0(\rho) =  \frac{c_0}{\rho} \nabla \rho
\end{equation}

This represents a situation where particles pile up
from the top to bottom with a certain density distribution.
For simplicity, the mean speed is assumed to satisfy eq.~\ref{drag}.
The mean speed as a function of density is plotted in Fig.1a.  
When the density of the granular particles exceeds the critical
density, say for example $\rho_c = 0.501$, 
the particles become locked and
have a steady-state density $\rho = \rho_c$.
If we solve the eq.~\ref{static} and 
plot the density profile as a function of z (Fig. 1b),
 we can see that
the uniform density profile is created close to the bottom of the
plate, and then the density decreases linearly to zero as we move to the
top of the container. Physically, this is acceptable if we notice that
when the particles neat the bottom
are locked, they act more like a solid.  
On the other hand, those particles near the surface
are rather loosely packed and 
ready to move, which is more like a traditional fluid, where
the density linearly decreases as the altitude increases. Note that the
granular bed is not subjected to 
vibrations at this time and thus this is a
fixed bed solution. Furthermore, the total number of particles
is conserved by the mass conservation law of Eq.\ref{mass}.
We emphasize that the non-slip fluid boundary conditions have been imposed
in solving the equations numerically, namely,
$v_{\bot} =0$ and  $v_{\|} = 0$ [8]

\centerline{\hbox{
\psfig{figure=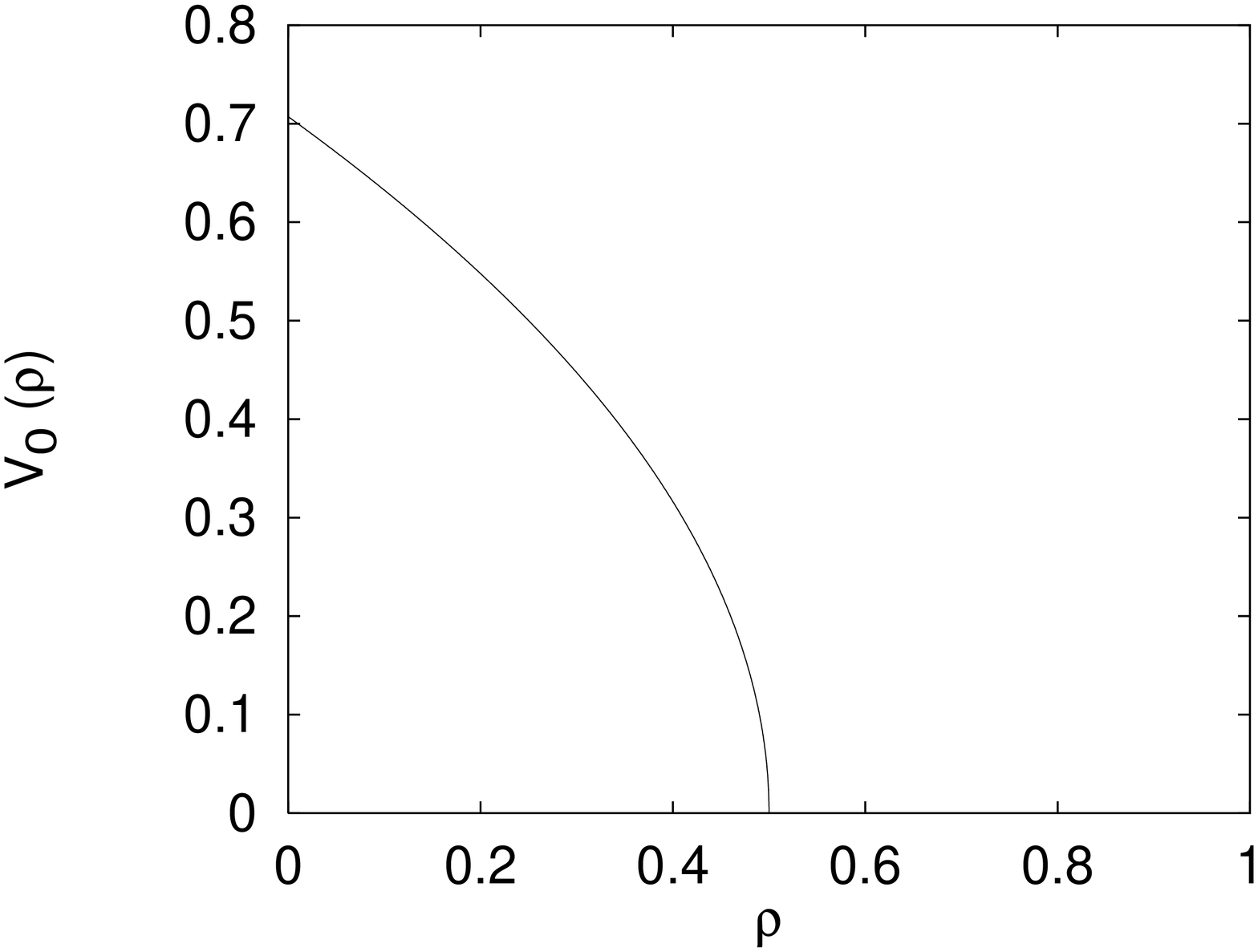,height=3.5in,width=3.5in,angle=270}
\psfig{figure=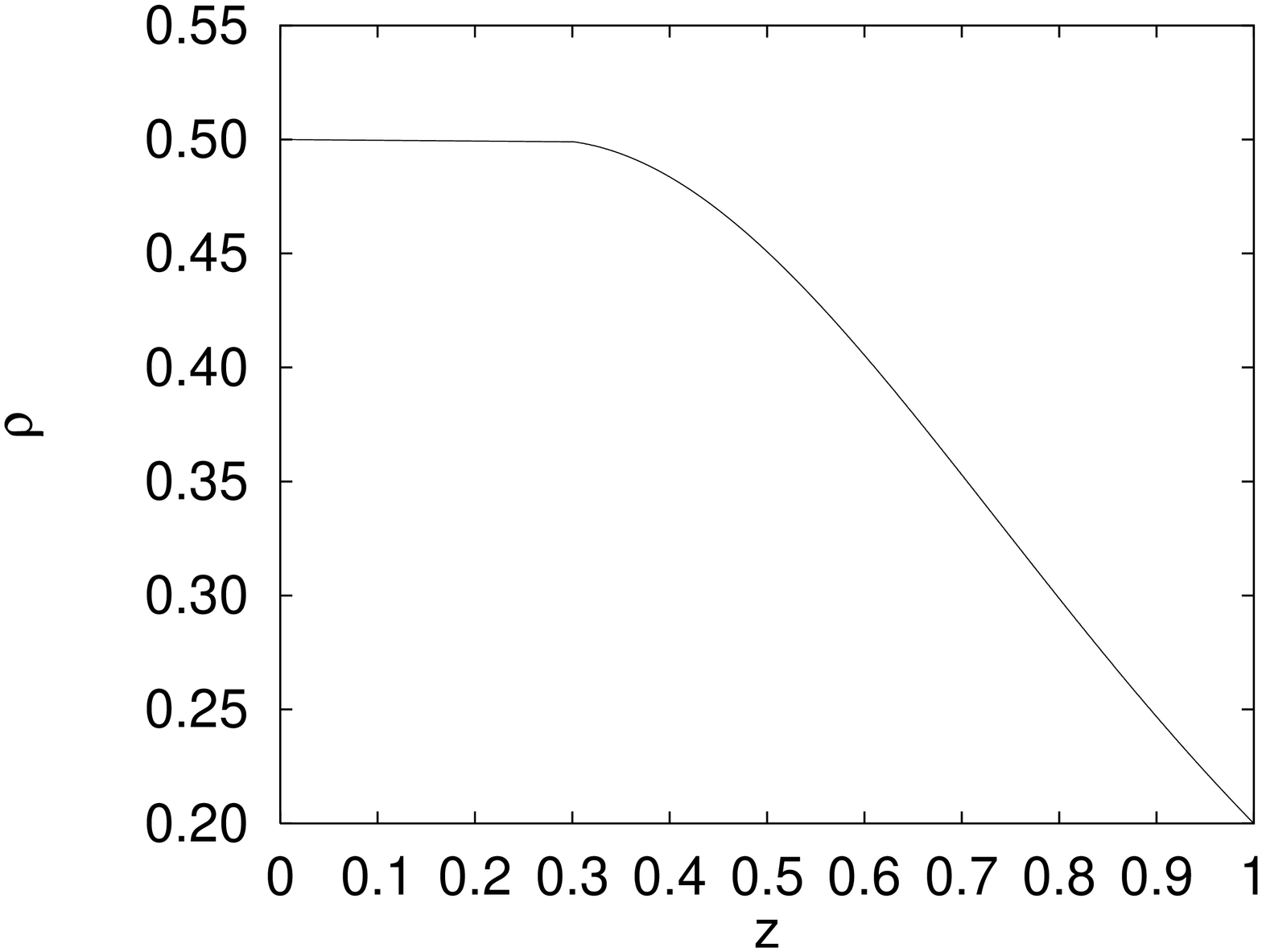,height=3.5in,width=3.5in,angle=270}
}}
\centerline{(a) \hspace{1.0in} (b)}

\noindent Fig.1. (a) Mean speed $V_0(\rho)$ as a function of density is plotted
above(a), with the  cutoff density $\rho_c = 0.501$ and $\beta=2/3$.
(b) Static solution is solved analytically and the density, $\rho$, is plotted 
as  a function of the height $z$, where $z = 0$ is the bottom of the container.
The  numerical result  is consistent with this, to a very high accuracy.
\vskip 0.5 true cm
Second, a spatially uniform solution.
There is, however, a second class of solution that exists only
when $V(\rho)$ has a cut off.  This solution is a spatially uniform solution
with the density given by the closed packed density, $\rho_c$, everywhere
with $V(\rho_c)=0$.  Numerical simulations of the traffic equations
reveal that it is the spatially nonuniform solution that is realized in
simulations.  But the stability analysis of the uniform solution clearly
shows that this solution is stable, too, and thus it must be realized with
presumably suitable boundary conditions.
Further, deep inside the bed, there is very little difference between the
uniform and nonuniform solutions, and thus the convective instability inside
the bulk may be investigated by focusing on the stability of the uniform
bouncing solutions.  In the absence of the rigorous stability analysis of
the nonuniform solution, this is only our guess.
Future work must focus on the rigorous stability analysis of
the nonuniform solution(see Appendix B) and compare its result to those
of the uniform solution obtained in section III.  
Before presenting the stability analysis, we first present numerical results.

\vskip 0.5 true cm
{\bf B. Numerical results}:
We now investigate the time dependent phenomena numerically.
Numerical solutions are obtained by two different algorithms for the 
rectangular boundaries. The first one is the center-difference
method, which has the accuracy of 
the second-order in space and first order in time. The second method 
employed here is the RUNGE-KUTTA algorithm which has the accuracy of 
the second-order 
in space and third order in time. The results show that both methods give 
consistent  conclusions for the same boundary conditions.
For most of the cases, we set the grid number equal to  $N_x=32$ and
$N_z= 32$ in order to improve the accuracy while accommodating 
the CPU cost. Lower or higher
grid numbers have been examined as well, but 
no significant changes have been found. Initially, the uniform solution 
is assumed, and typically four cycles are taken to calculate the
average of the convection after the 
stable periodical solution appears at each site in the lattice.
In the stream line plot, 
all of the vector flow arrows are normalized by the maximum speed.
Numerically, we observe two distinctive classes of solutions: the first
one is the bouncing solutions, and the second one the convective rolls.
We first examine the bouncing solutions.  Note that
the set of parameters that correspond to physical situations might be
$\omega/2\pi=20Hz$, $T_e\sim 3$, and
$\mu=1/R \approx 0.5$ with the Reynolds number $R=UL/\mu_s \approx 2$,
because the kinetic viscosity of the granular material, $\mu_s\approx 5\bullet
10^{-3} m^2/s$ and the typical velocity $U\approx 10 cm/s$ [8].  For pure
numerical reasons, however, in what follows, unless mentioned
otherwise, all the simulations
are conducted with the parameters, $\mu=1,\omega=1,\tau=1$ and
$T_e=2$, L=32 with $V_o(\rho)$ given by Fig.1.  

{\bf (a). Bouncing solutions with $\Gamma$ $\leq$ $\Gamma_c$}:
For a fixed set of control parameters,
bouncing solutions, first termed in [8],
appear when the vibration strength $\Gamma$ is less
than the numerically determine critical $\Gamma_c \approx 1.5$.  
Upon increasing
$\Gamma$, this bouncing solution becomes unstable and bifurcates into
a different branch and convective rolls appear.
We have examined the case where the vibrational acceleration 
is less than the gravity, in other words, 
the dimensionless variable $\Gamma$ is less than one, which is
certainly the regime below the critical strength.  In real situation, for 
$\Gamma < 1$, nothing much 
really happens inside the box except perhaps the volume
decreases due to settling caused by vibration. However, in our simulation,
due to the presence of the top layer, we expect the bouncing solution to 
appear even for $\Gamma < 1$.  We have performed the simulation
using different initial conditions such as a uniform density distribution, 
a linear density distribution, or a fluctuation in the velocity field. 
For all cases, we have found that after 
a relatively long CPU time, the convection pattern disappears,
which implies that 
the convection does not persist in this parameter regime.
However, if we
measure the local density or velocity at different locations, we clearly
observe oscillatory solutions that
persist everywhere inside the container.
This is termed the ``bouncing solution'' that exists
before the onset of the convection:
the bulk of the granular material oscillates up and down, like a single
solid ball, with a central frequency of the vibrating container and some
harmonics. The harmonic oscillations of the density and
velocity at the same lattice site are displayed in Fig.2a and 2b.
\vskip 1.0 true cm
\centerline{\hbox{
\psfig{figure=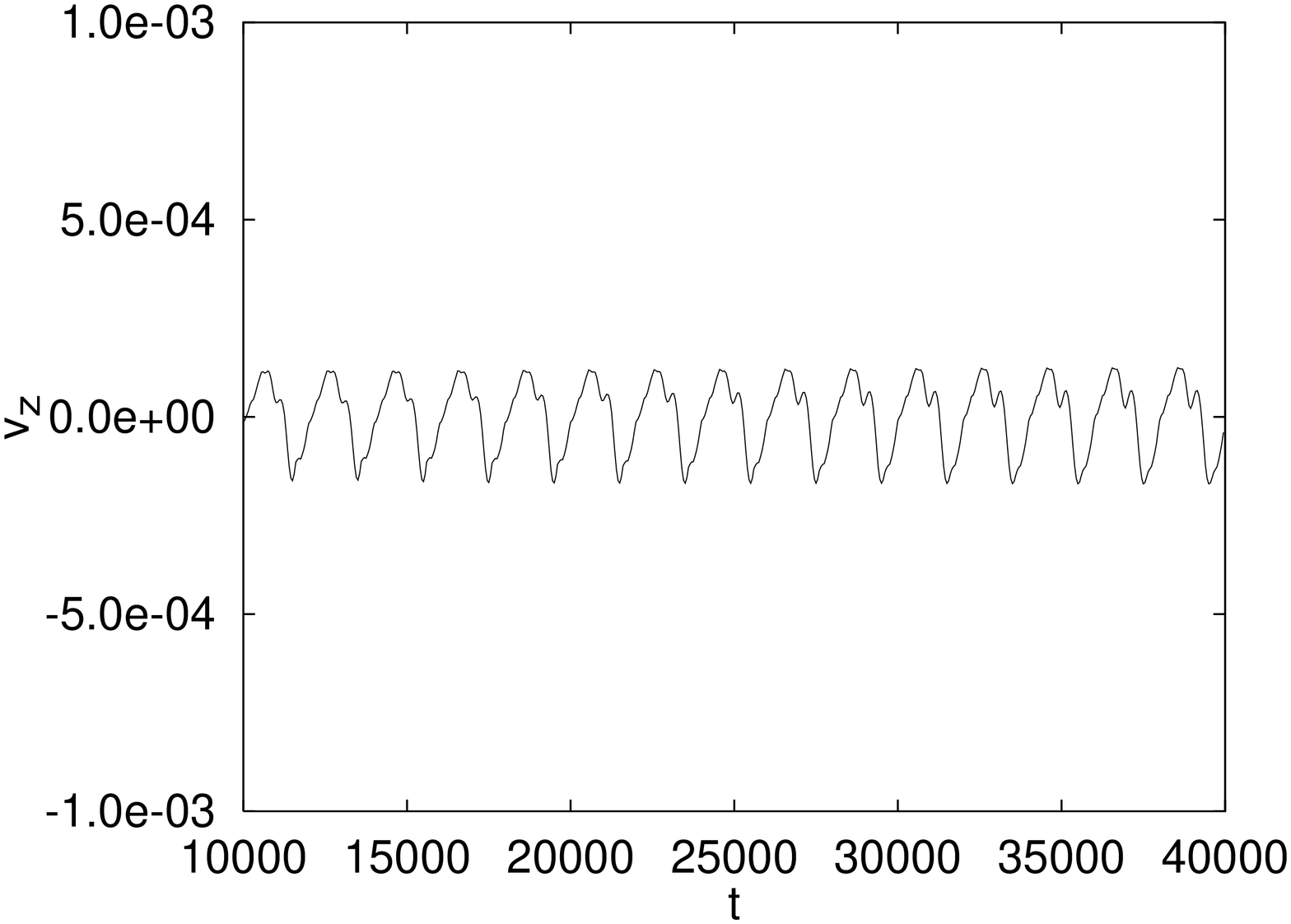,height=3.5in,width=3.5in,angle=270}
\psfig{figure=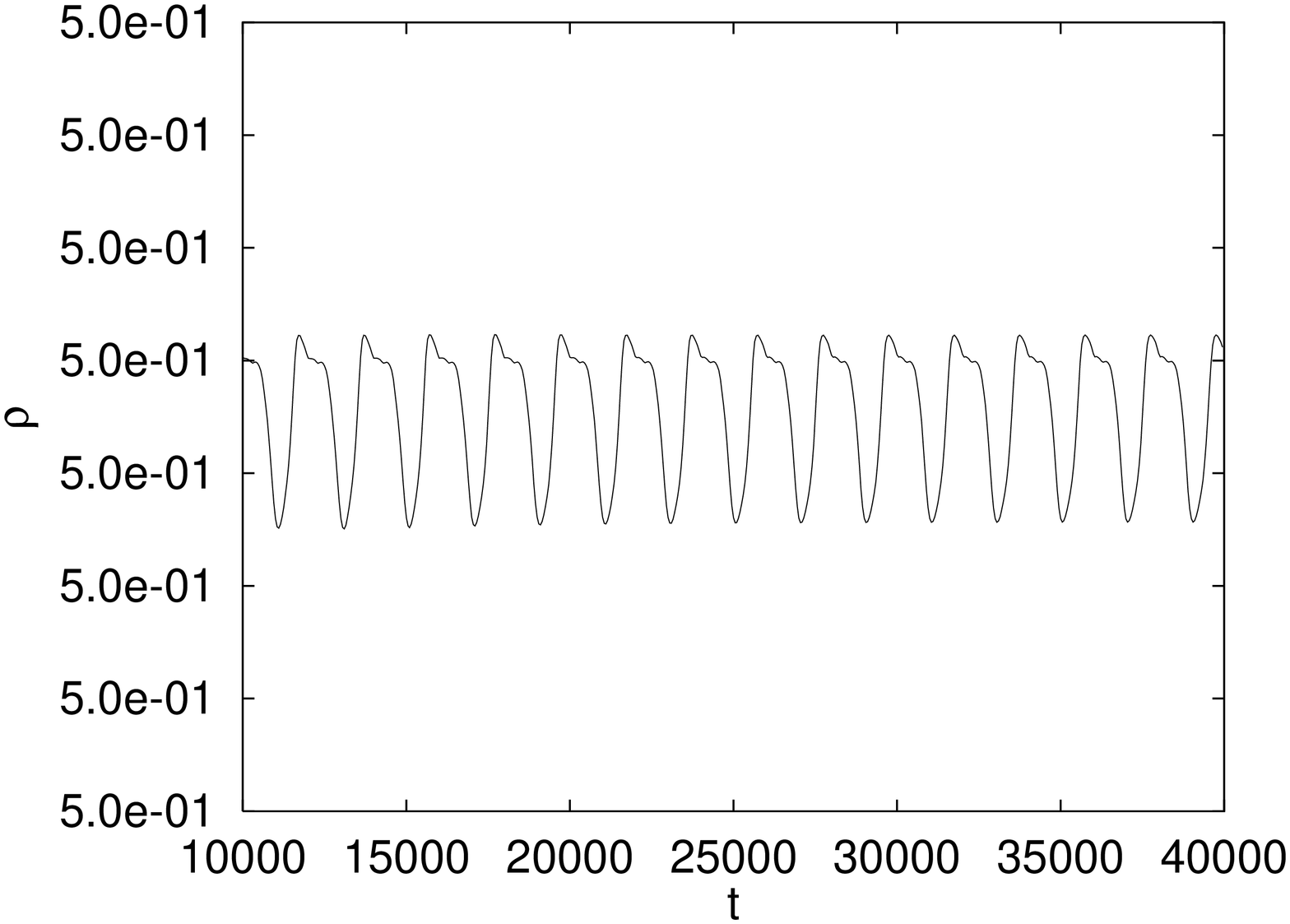,height=3.5in,width=3.5in,angle=270}
}}

\noindent Fig.2. Bouncing solutions for (a) 
the velocity $v_z$ and (b) the density at
the same fixed location (x,z)=(15,15), where no convection rolls appear
and the bulk of the particles moves as a single particle.  $\Gamma <\Gamma_c$.

\vskip 1.0 true cm
{\bf (b). Convective domain with $\Gamma > \Gamma_c$}:
When $\Gamma > \Gamma_c$, the bouncing solutions disappear and the permanent 
convective rolls appear inside the bulk.
In order to view this slow convection,
we have to take the average of the velocity field over many periods.
Typically four cycles are used, 
since the amplitude of the convection is relatively small in 
comparison to the amplitude of the vibration.  This is true in experimental
systems as well.
In the stream-line plot, we have normalized the velocity field,
by taking the maximum velocity component as the maximum length of the arrows.

In Fig.3a convection patterns are illustrated
for $\Gamma = 2.25 >\Gamma_c $
by plotting the streamline of the velocity field, where
$V_0 = (\rho_c-\rho)^{\beta}\theta(\rho_c-\rho) $ with $\beta=2/3$ and
$\rho_c=0.501$ as given in Fig.1,
is used. Notice that the 
flow direction close to the wall is toward the upward direction, and 
at the center of the container, the particles try to move up. The 
density profile is also plotted in Fig.3b, 
where the average of the density over
one period remains close to the static solution. At any given time the
density profile varies around such an average velocity.

\centerline{\hbox{\hspace{0.4in}
\psfig{figure=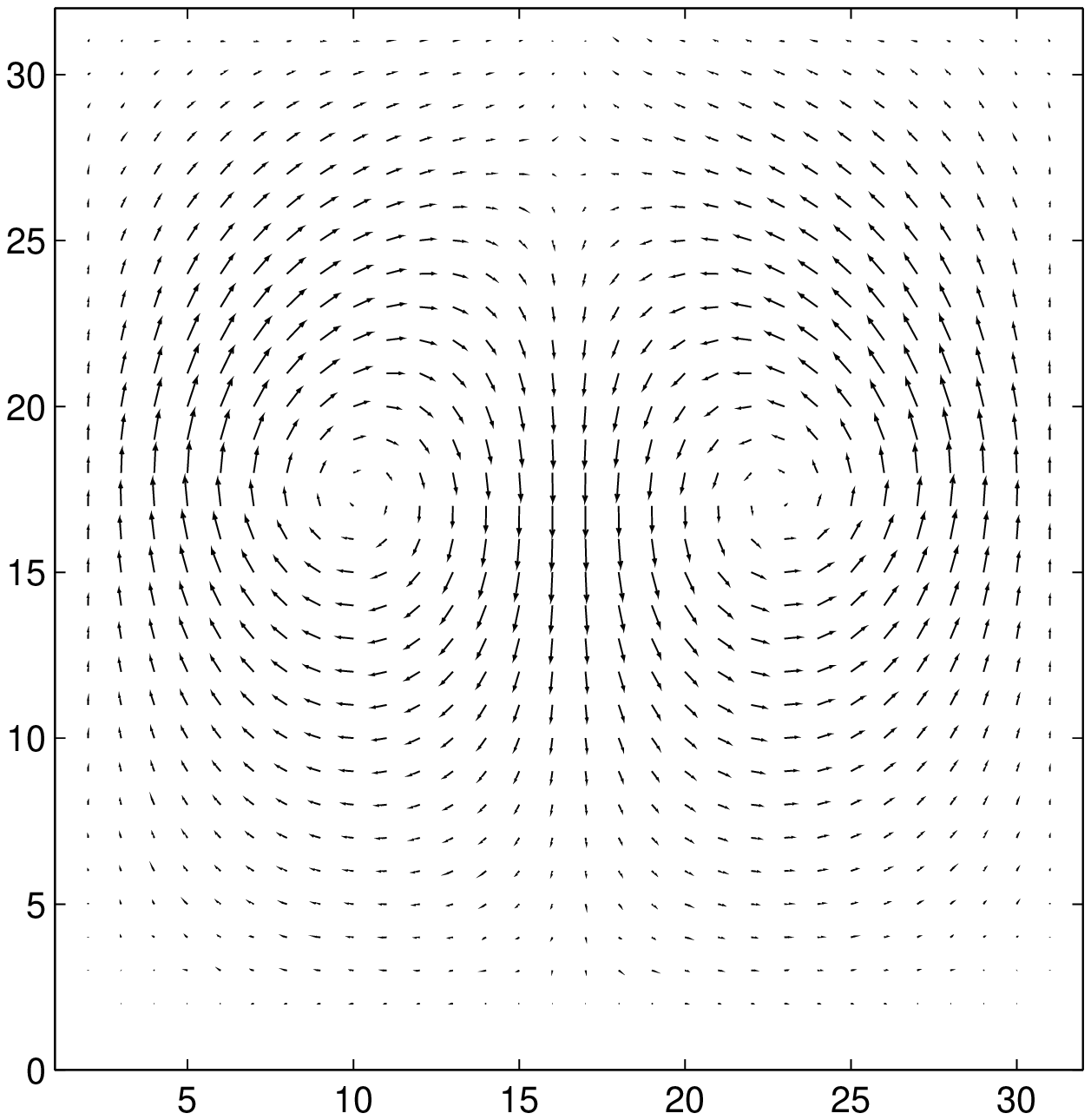,height=3.0in,width=3.5in}
}}
\vspace{.3in}
\nopagebreak
\centerline{\hbox{\hspace{-0.3in}
\psfig{figure=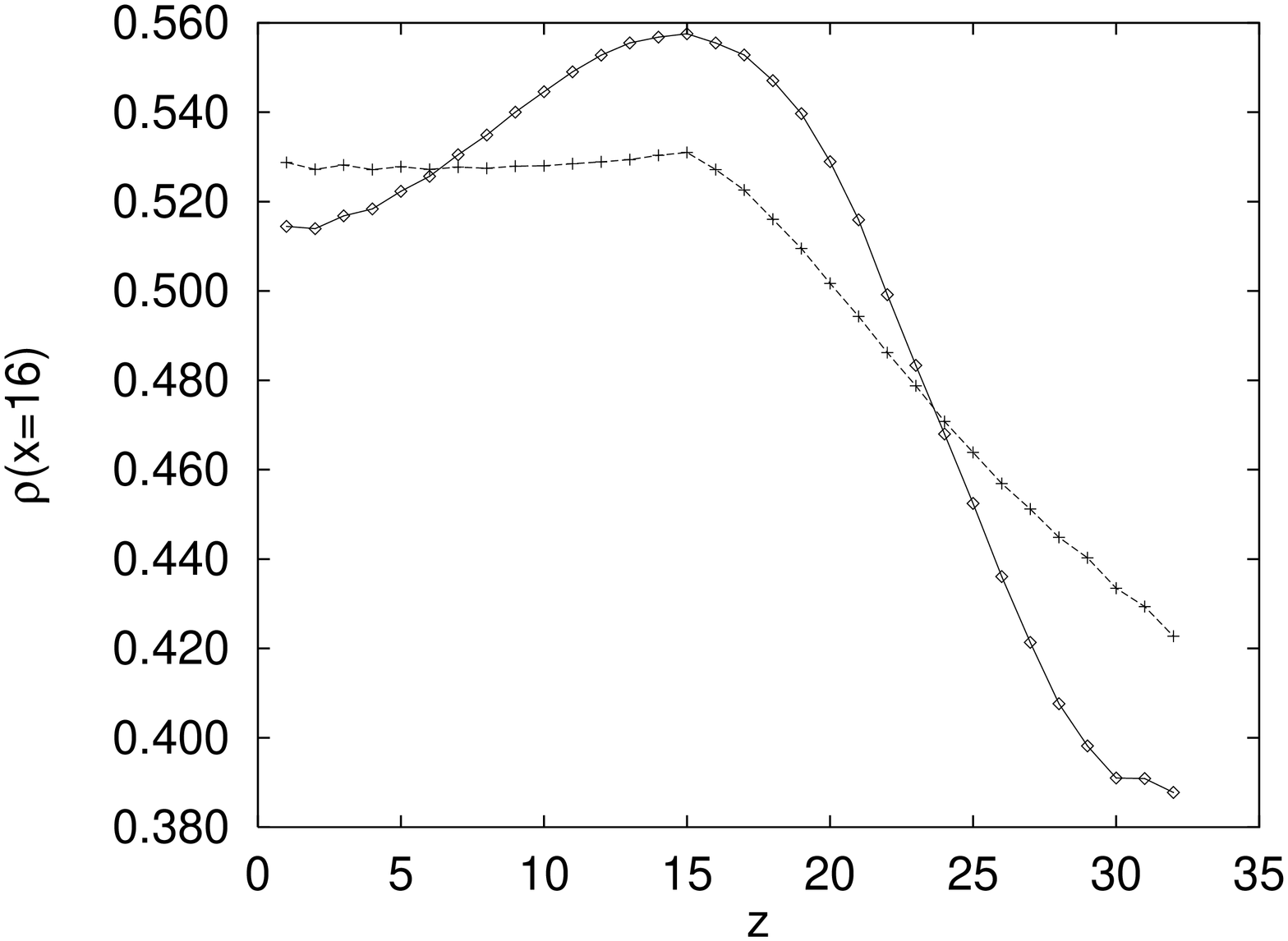,height=2.0in,width=3.5in,angle=270}
}}
\vspace{-0.2in}

Fig.3. Convection pattern and density profile:
(a) When $\Gamma > \Gamma_c$, convection patterns are illustrated by
plotting the streamline of the velocity field, where the 
velocity field is normalized by the maximum velocity component $v_z$,
with $V_0$ given by Fig.1.
(b) The
density profile as a function of position.  The
average of the density over
one period remains close to the static solution(+) and at any given time the
density profile(circle) swings around such an average velocity.

\vskip 0.5 true cm
Next, 
it has been found that the geometry plays an important role in the convection
pattern formation [7,22] and we examine
its effect on the convective instabilities. 
For example, the change of the aspect ratio(defined as width/height)
may affect 
the relative positiona, the number of convection rolls, and the intensity of 
the convection with respect to the same vibration parameters.
We have examined two particluar cases:

First, the numerical results 
displayed in Fig.4a and 4b show that the location of the convection rolls
migrate when the aspect ratio changes.  For the aspect ratio, $Nx/Nz=0.5$,
convection rolls move toward the top of the container, 
and the particles close
to the bottom are almost locked allowing no relative motion to exist.  This
is consistent with the experiments and MD simulation results.
Convection rolls are shown in Fig.4a and the corresponding density profile
is plotted in Fig.4b.  Note that
the stability analysis of model A[8] predicts 
a series of rolls when the vertical axis increases as a multiple of
the horizontal width.  The form of $V(\rho)$ with a cut off at the
closed packed density seems to be crucial in capturing the real convective
patterns.  Further,
the average density curve ($+$) is very close
to the static solution without the vibrations. The second curve ($-$)
is the density profile at the stop time $t$, which clearly shows that the
density profile varies in time around the equilibrium density profile.

\centerline{\hbox{\hspace{0.6in}
\psfig{figure=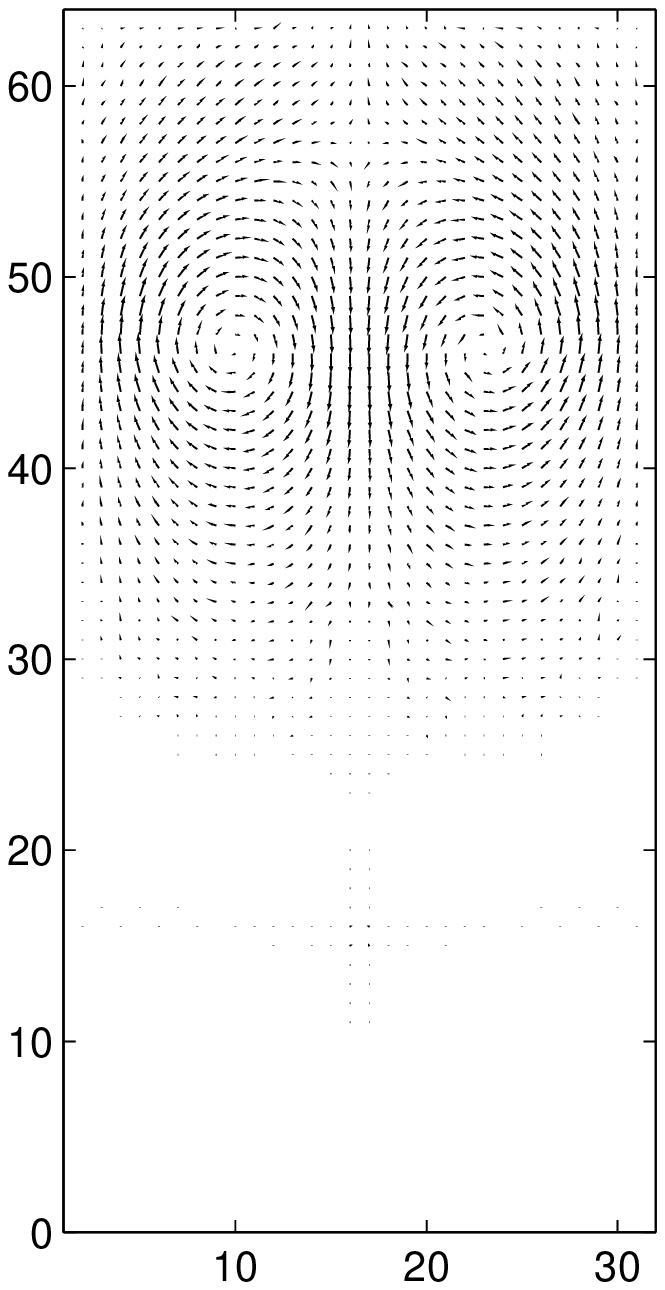,height=3.0in,width=2.5in}
}}
\vspace{.3in}
\nopagebreak

\centerline{\hbox{\hspace{-0.3in}
\psfig{figure=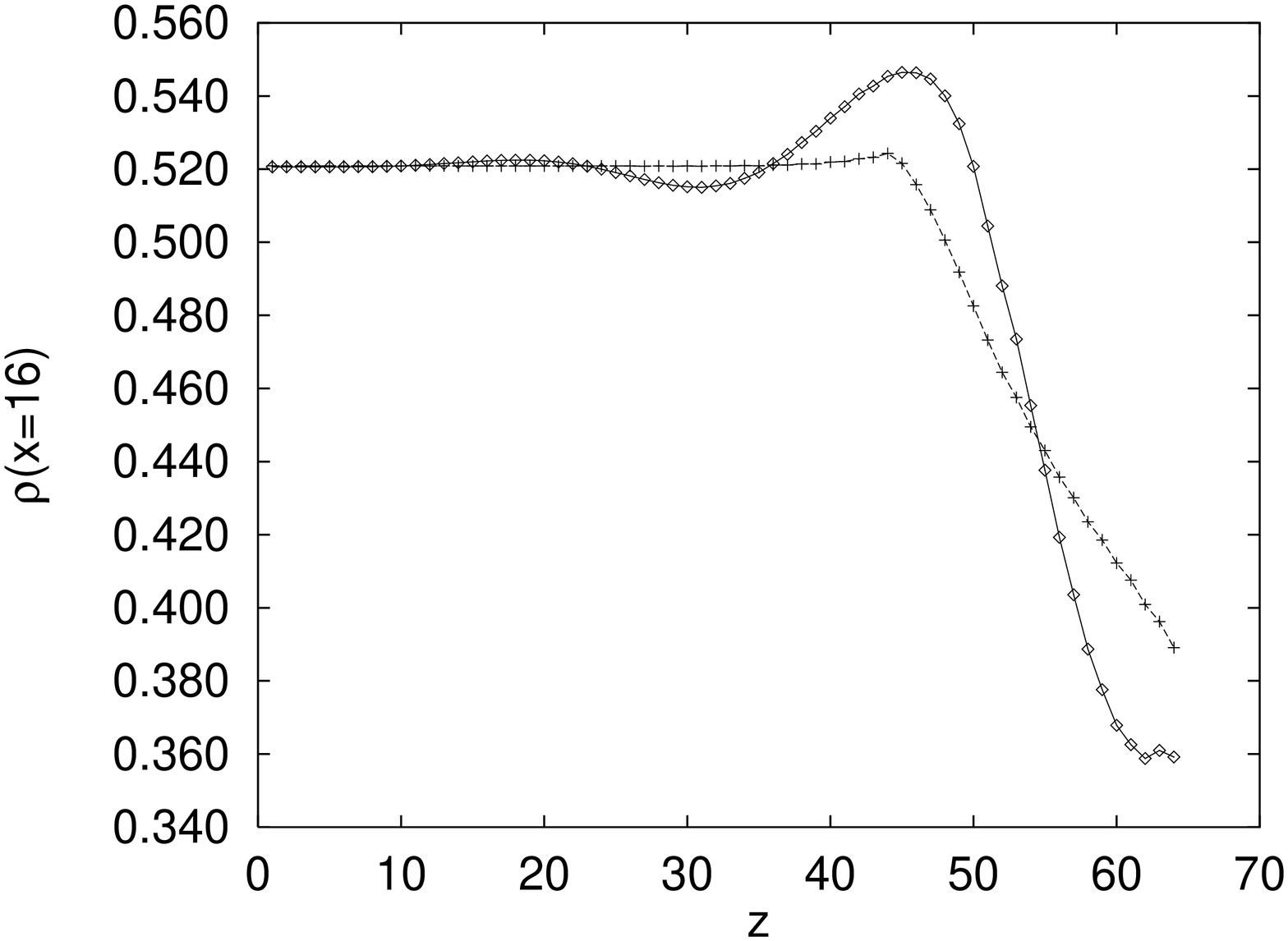,height=2.0in,width=3.5in,angle=270}
}}
\vspace{-0.2in}

Fig.4. Convection pattern for a small aspect ratio
$N_x/N_z$ = 0.5.
All other parameters remain the
same as in the case of
$N_x/N_z$ = 1.0. (a) The two convection rolls move to the
top region of the container and most of the particles closed to
the bottom are locked during the whole vibration cycles. This can be verified
in the density profile at a fixed point x=16 (b), where the time dependent 
density curve(circle) varies around the
static solution(cross), but much less at the bottom than at the top regime.

\vskip 0.5 true cm
Second, we have also examined the case with the large aspect ratio, 
$N_x/N_z=2.0$, (results are
shown in Fig.5a), where the width is twice larger than
the height. Here, four
convection rolls appear with two small ones near the top 
corners (Fig.5a).
If we extend the width even longer to the aspect ratio, $N_x/N_z=4$,
the convection pattern will 
only exist around the narrow regime of side-boundaries,
represented by the zero intensity of convection (Fig.5b).  
It seems that convection rolls only appear close to the
boundaries, which indicates that the friction on the wall 
presumably plays a significant
role in the origin of the vibration-induced convections.  Such observations
are consistent with the observations
made by Taguchi in Molecular Dynamics simulations [22].

\vspace{.3in}
\par
\centerline{\hbox{
\psfig{figure=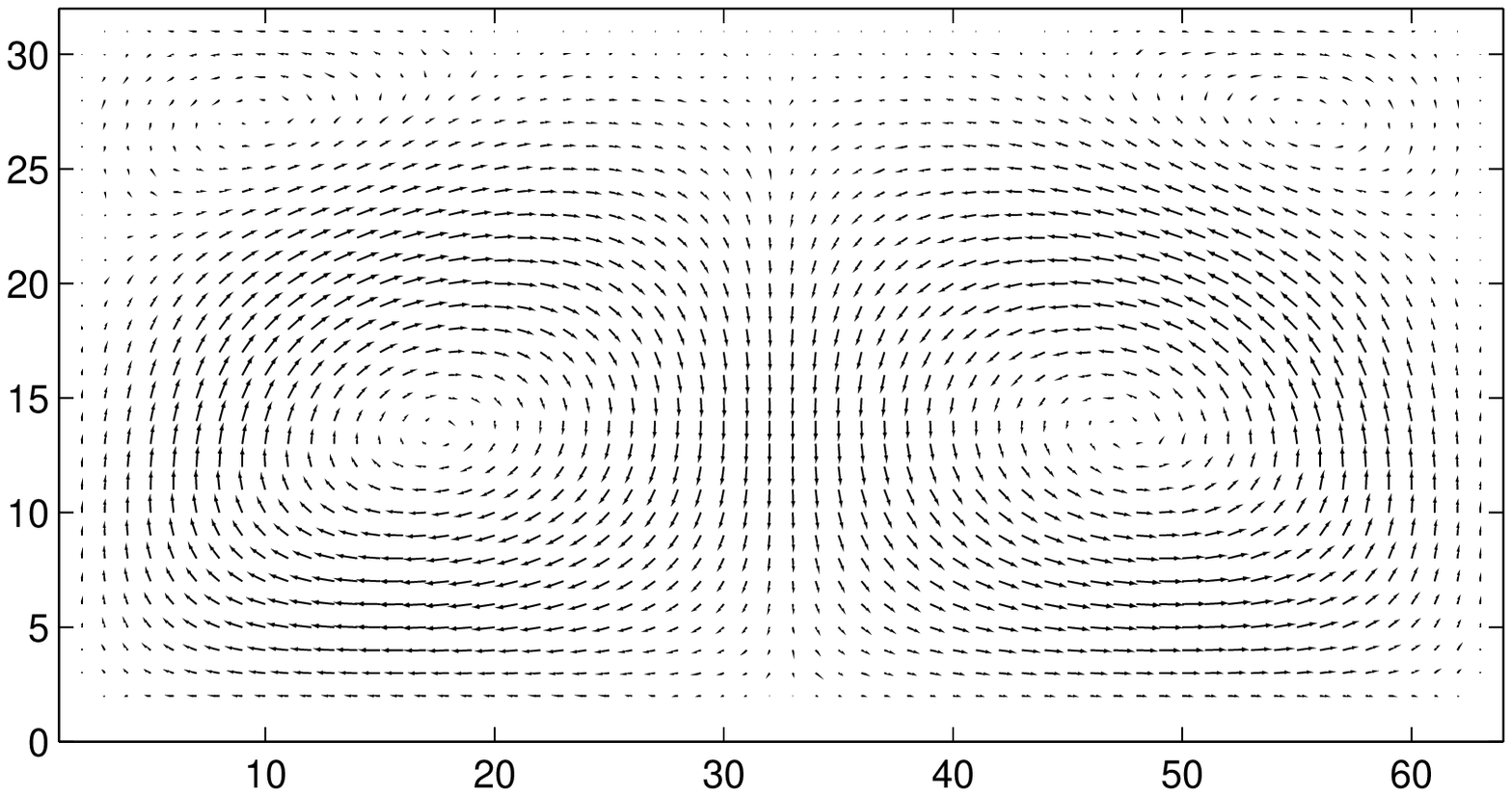,height=2.0in,width=4.0in}
}}
\vspace{.3in}
\nopagebreak
\centerline{\hbox{
\psfig{figure=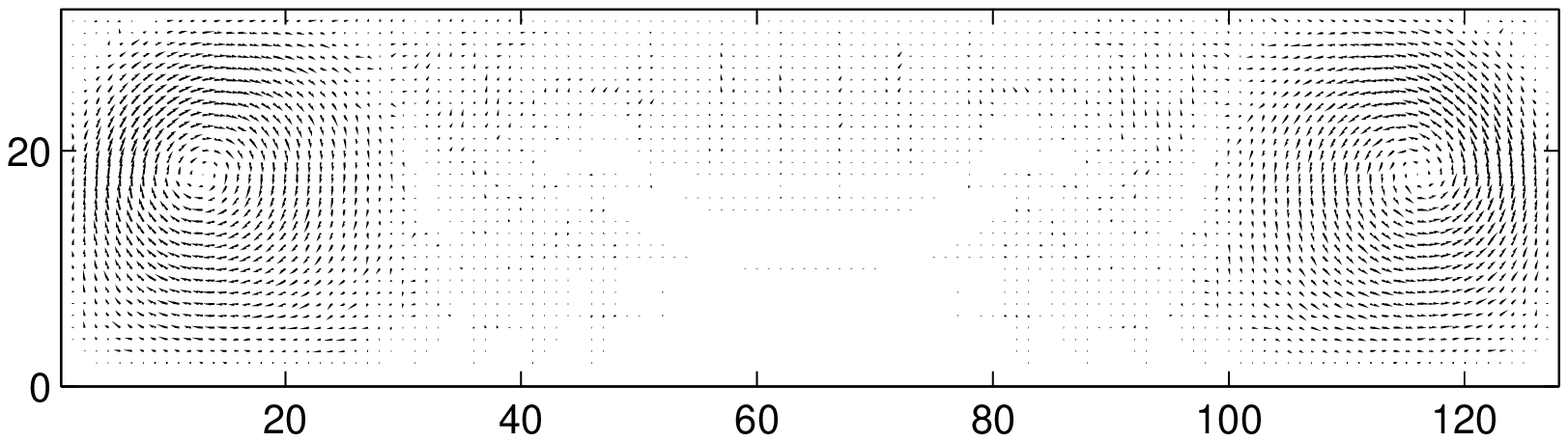,height=2.0in,width=7.0in}
}}
\vspace{-0.2in}

Fig.5. Convection pattern for large geometry ratio, 
(a) Convection pattern is drawn for a large aspect
tratio, $N_x/N_z$ = 2.0.  All other parameters remains the
same as in the case of
$N_x/N_z$ = 1.0. Two main convection rolls appear at the 
center, with another two visible small convection rolls at the top corners
of the container. The two small convection rolls have opposite spins
relative to the major ones.
(b) Two convective rolls for
the aspect ratio $N_x/N_z = 4.0$ are illustrated  for the
same parameter regimes.  The two rolls  
exist  near the walls. They are separated by a wide non-convection
regime, where the velocity field equals zero.  

\noindent {\bf C. Pattern dependence on $V_0(\rho)$}
While we have not systematically examined the stability diagram of the
many dimensional parameter space, we report here a few cases simply
to show the richness of the problem associated with the traffic equations.
In the previous arguments, we have only assumed that $V_0(\rho)$ likely
has a cutoff when the density exceeds a critical density(at which 
the particles become locked),
 and reaches the maximum when the density approaches  zero (since
there are no interactions).
On the other hand,  it is also reasonable that  $V_0(\rho)$ might have
a long tail in the high density regime, which may represent  very rough
grains or grains non-uniform in size.  Hence, we have examined numerically
the case with the following $V_o(\rho)$.
$$V_0(\rho) = \alpha / (1 + \beta \rho^{2})$$
Our numerical results indicate
that the direction of the convection rolls is quite
different from the results obtained with a $V_0(\rho)$ that assumes a cutoff.
Moreover, the four roll convection pattern appears when the long-tail
$V_0(\rho)$ distribution is assumed. In Fig.6 is plotted
the convection streamline with $\alpha=0.2$,
$\beta =0.5$.
As we increase the vibrational frequency,
the high intensity regime of the convection rolls moves toward the boundary.
At some parameter regime, the two roll patterns evolve to four rolls.
In Fig.7, the four roll convection pattern is plotted.

In Fig.6, Fig.7, Fig.8,
 and Fig.9,
a wide parameter range has been examined.
It seems that the appearance of the two roll pattern or four roll pattern
not only depends on the aspect ratios, but also  on the 
vibrational frequency and amplitude.  Future work must focus on
exploring the parameter space as well as the effect of boundary conditions
to the convective patterns.
We now present the linear stability analysis of the
two dimensional traffic equations.
\vskip 0.5 true cm

\vspace{.3in}
\centerline{\hbox{
\psfig{figure=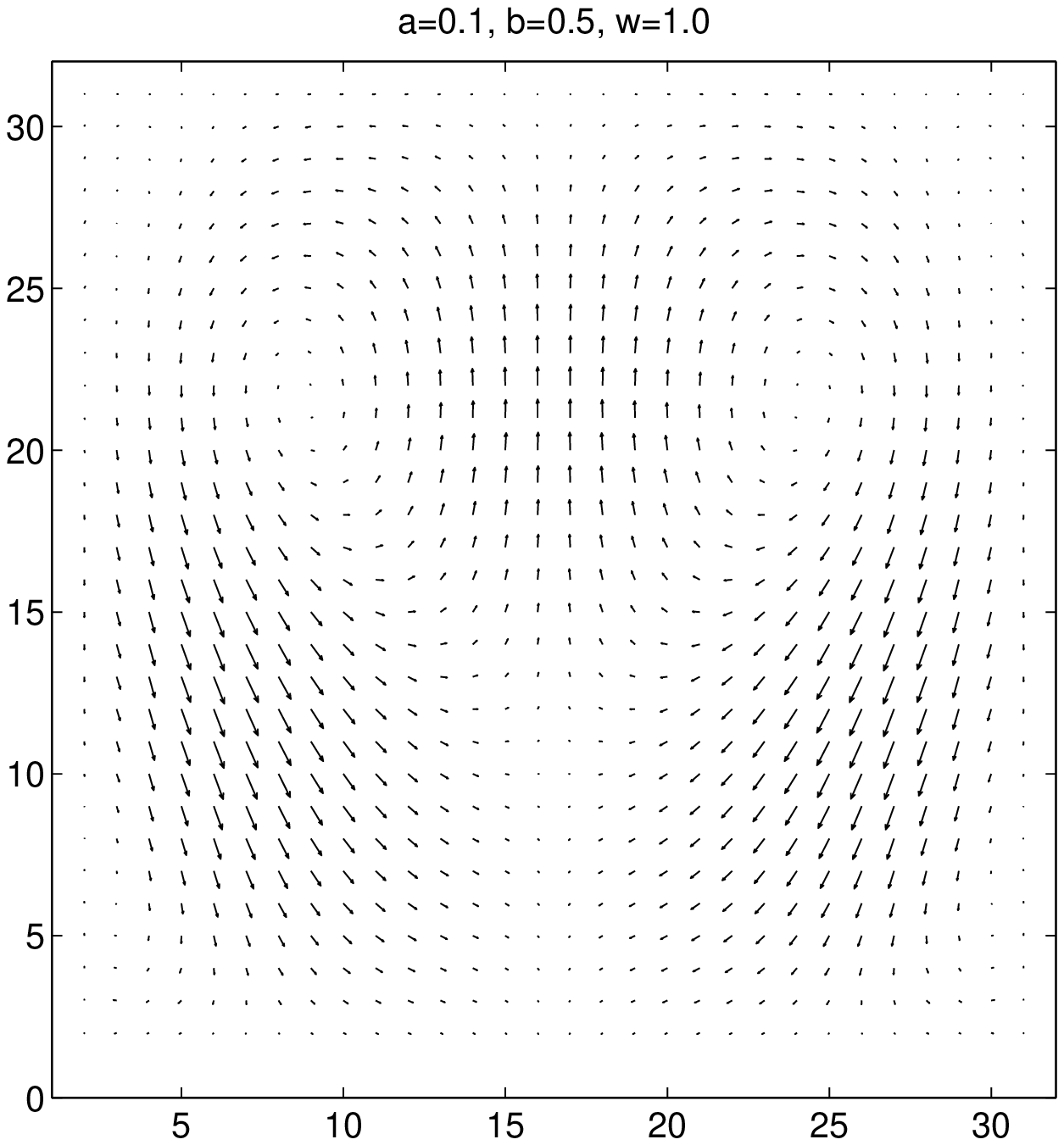,height=3.5in,width=3.5in}
\psfig{figure=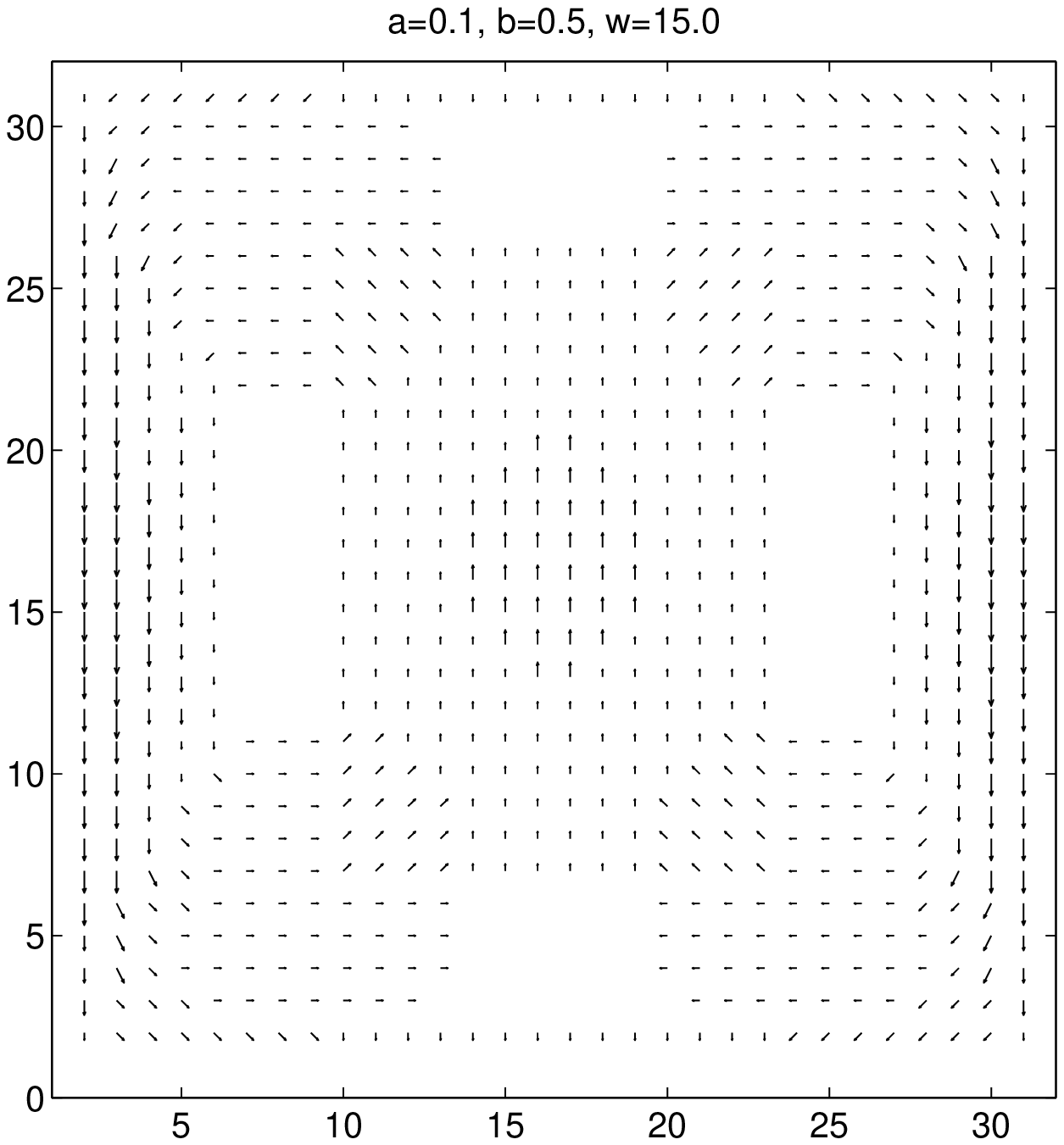,height=3.5in,width=3.5in}
}}

 Fig.6. Convection patterns for slow mean speed:
For slow mean speed $\alpha = 0.1$, vibrating 
amplitude $A = 2.0$, $\beta = 0.5$. Two convection rolls appears for the
different vibrating frequency, where higher frequency corresponds to 
a higher flow rate close to the boundaries, respectively from left to 
right $\omega = 1.0$ (a) and $\omega = 15.0$ (b).

\vspace{.3in}
\centerline{\hbox{
\psfig{figure=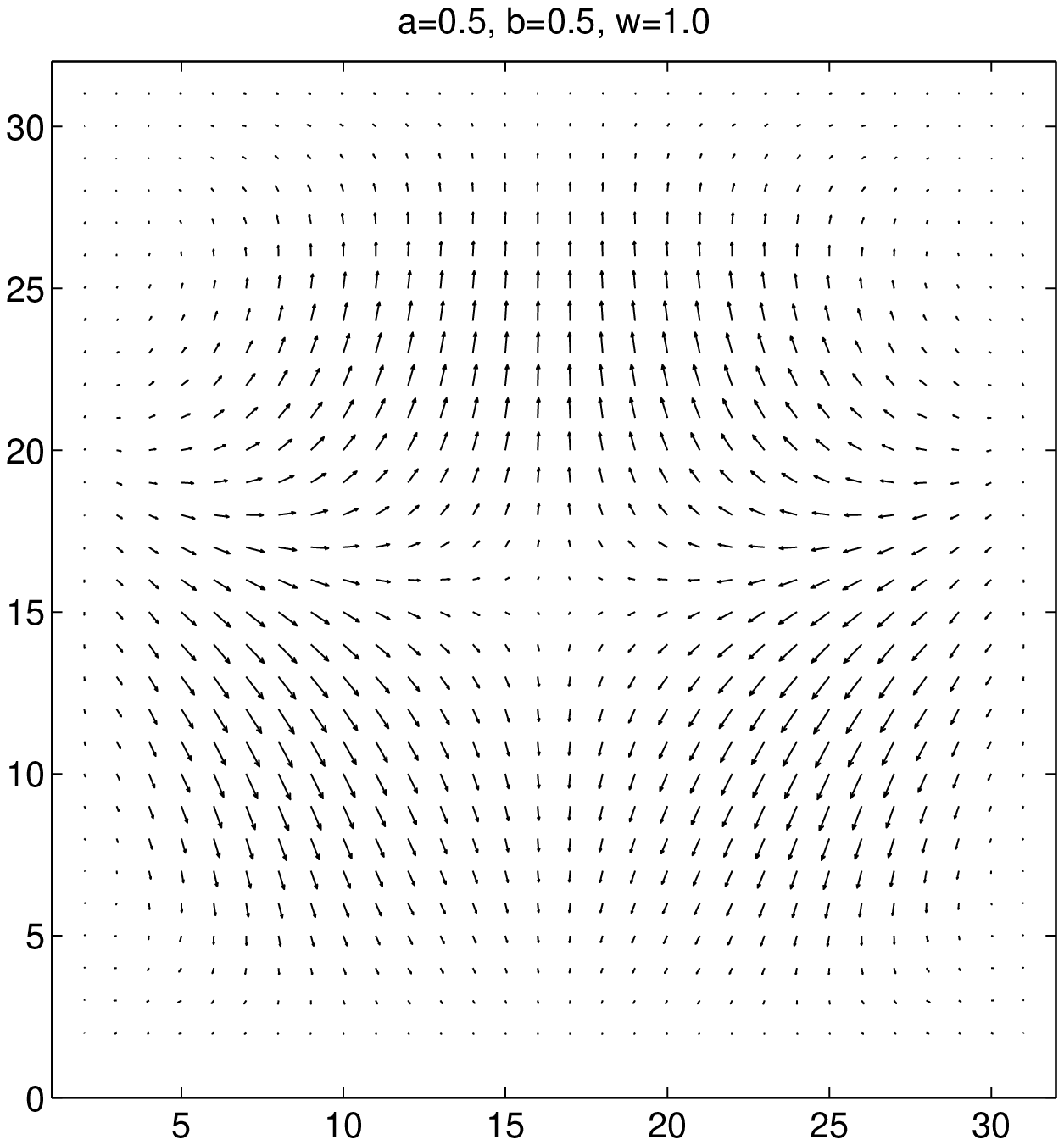,height=3.5in,width=3.5in}
\psfig{figure=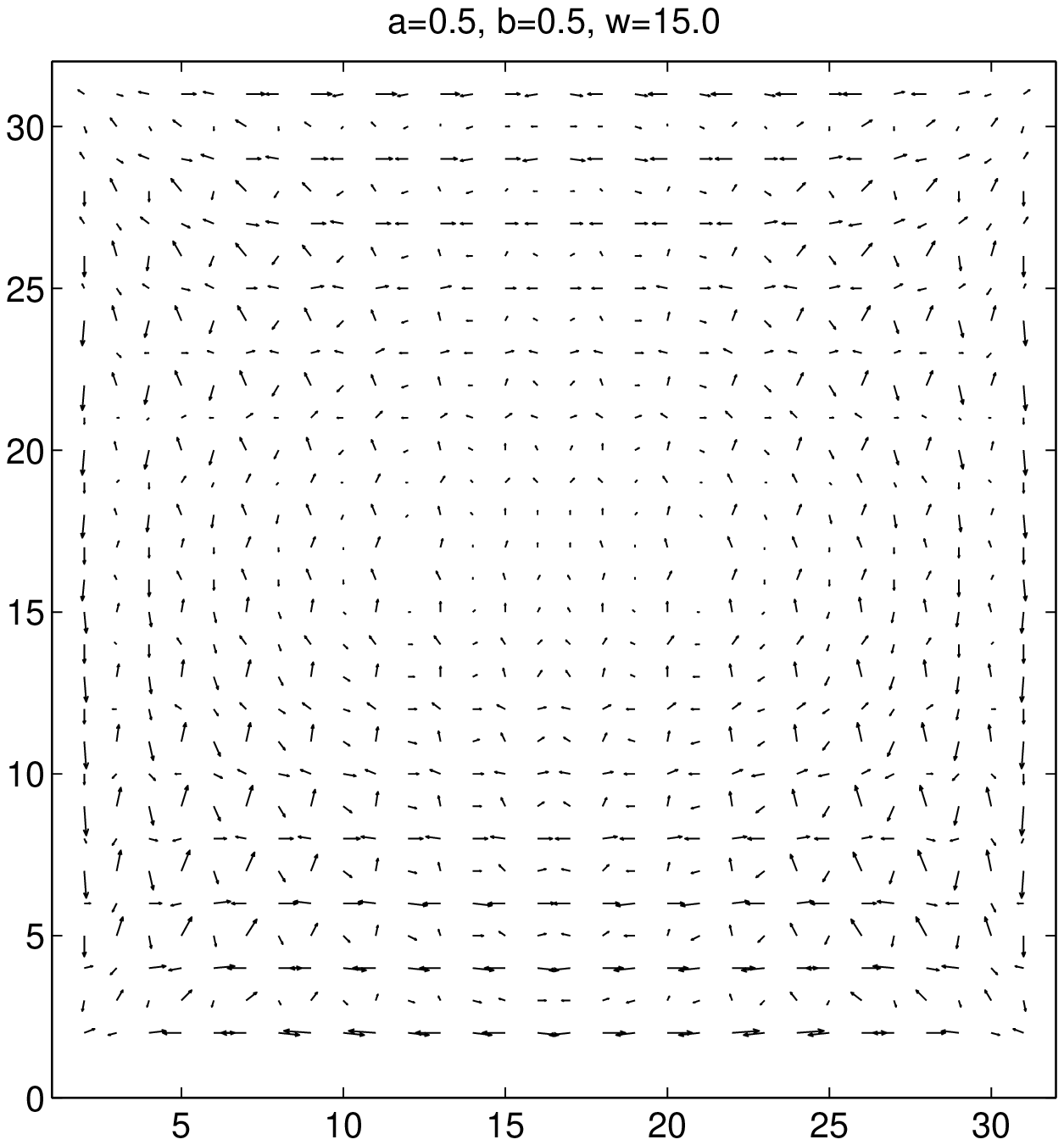,height=3.5in,width=3.5in}
}}

Fig.7 Convection pattern with four rolls: 
(a) Four convection roll appears when we increase 
the amplitude of the mean speed $\alpha = 0.5$ with the frequency
 $\omega = 1.0$. (b) 
As we increase the vibrating frequency, two roll convection
patterns appears with a higher flow rate close to the boundaries.

\vspace{.3in}
\par
\centerline{\hbox{
\psfig{figure=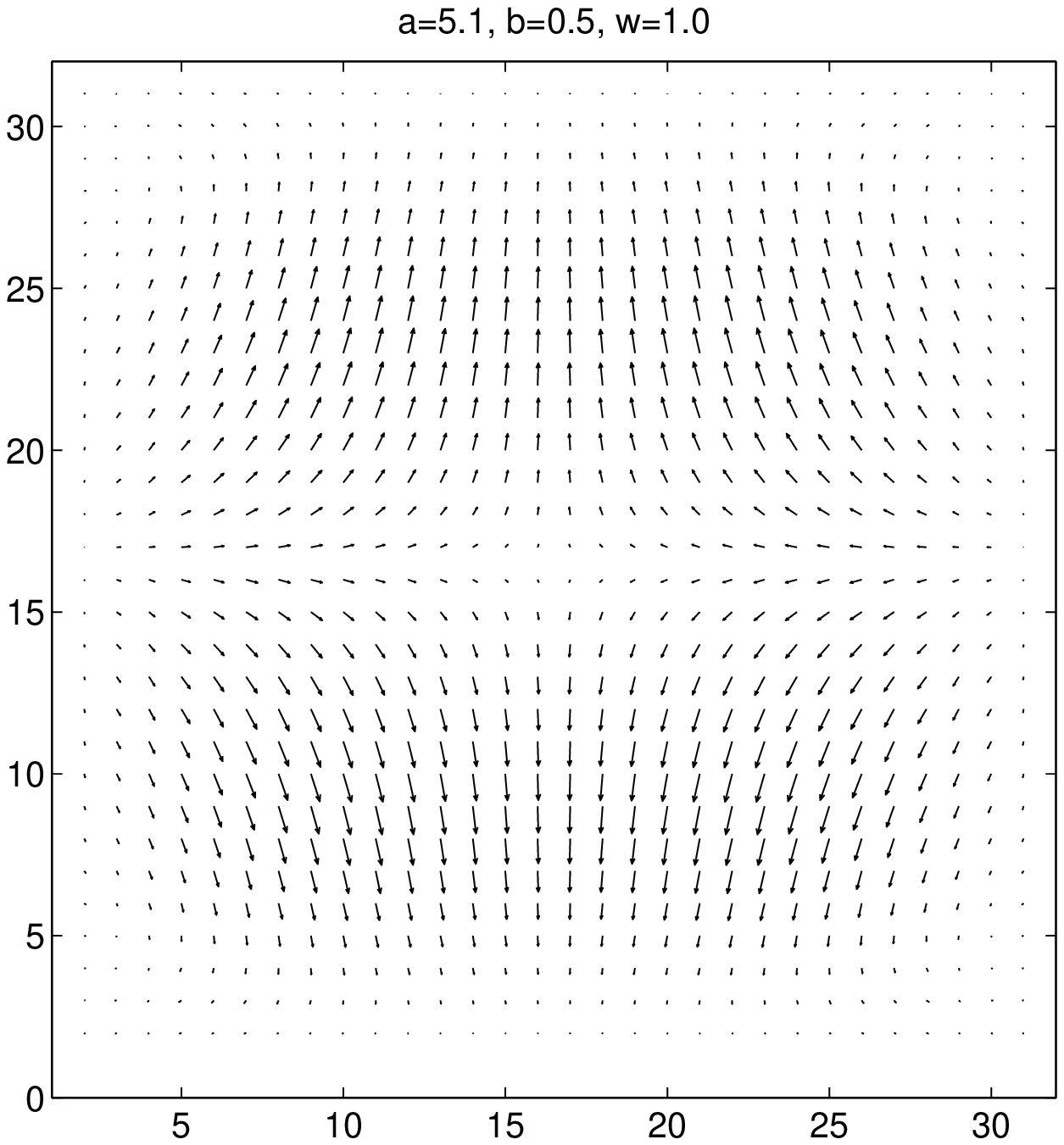,height=3.5in,width=3.5in}
\psfig{figure=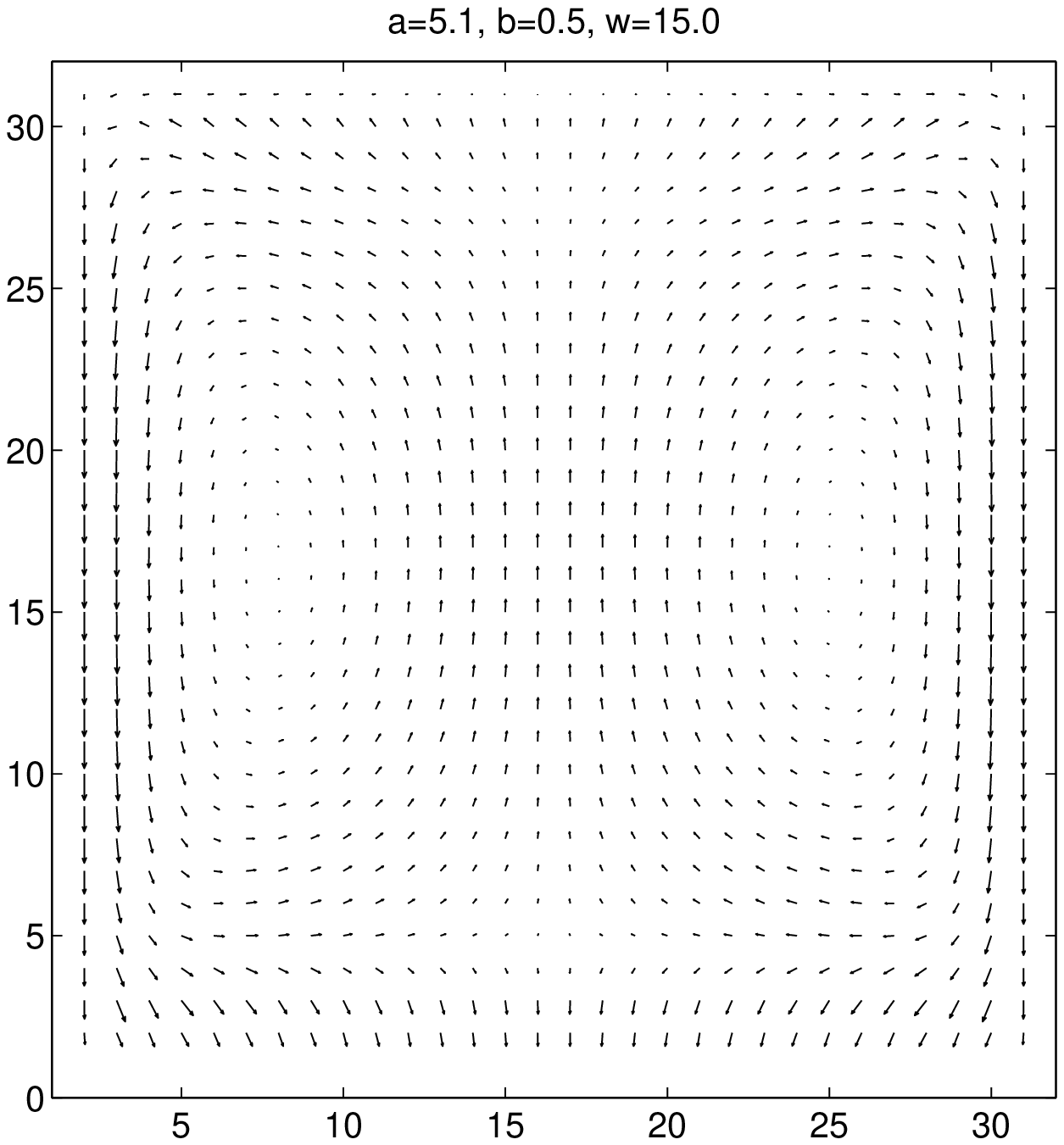,height=3.5in,width=3.5in}
}}

Fig.8 Convection pattern with four rolls:
(a) Four convection roll persists when we increase 
the amplitude of the mean speed $\alpha = 5.1$ with the frequency
 $\omega = 1.0$. (b) 
As we increase the vibrating frequency, two roll convection
pattern appears with a higher flow rate close to the boundaries.

\vspace{.3in}
\par
\centerline{\hbox{
\psfig{figure=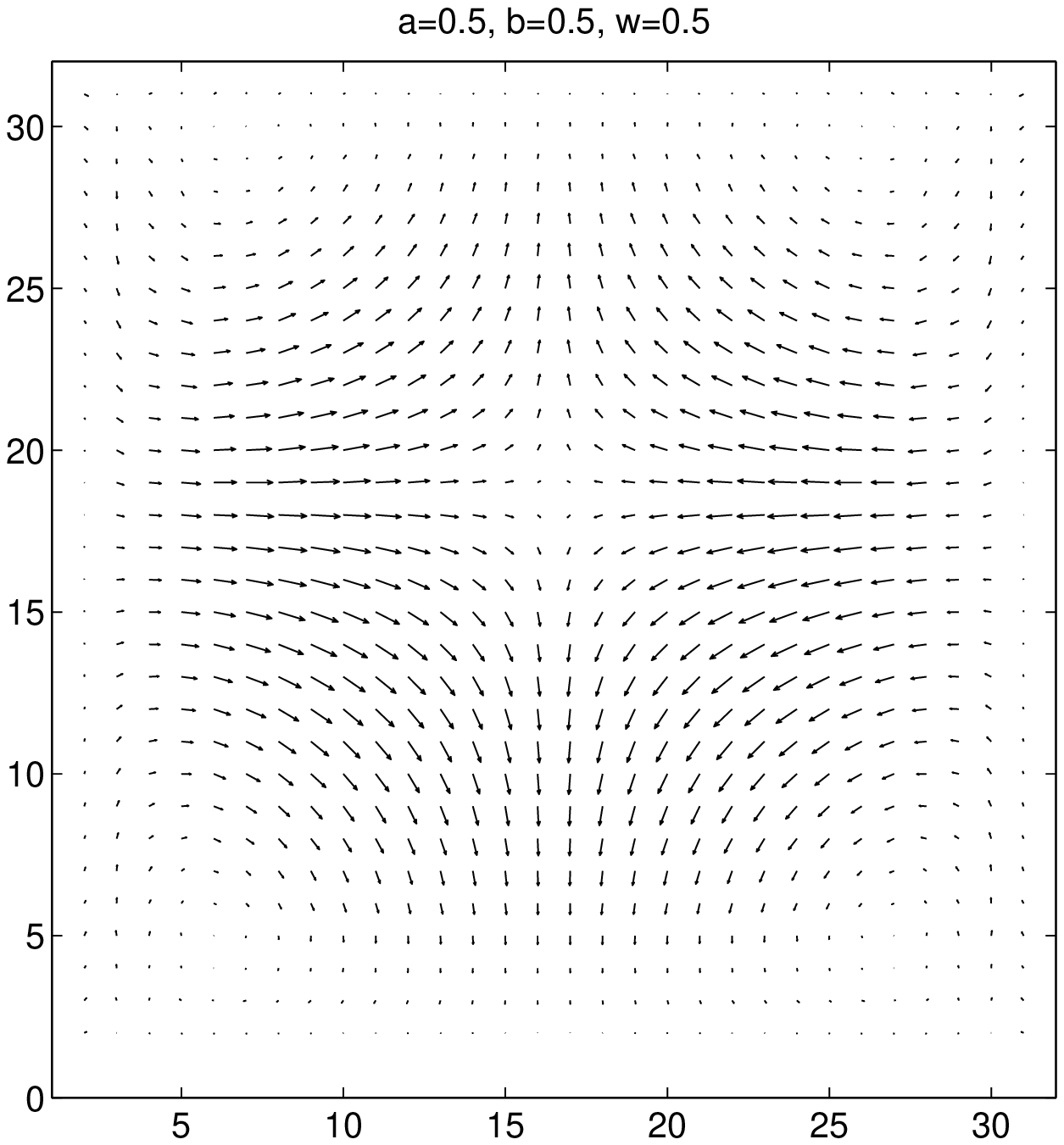,height=3.5in,width=3.5in}
\psfig{figure=a05b05w150.ps,height=3.5in,width=3.5in}
}}

Fig.9. Convection patterns with high frequency:
(a) For the same $\Gamma = 2.0$ as in Fig.8, which corresponds to 
the induced vibrating gravity acceleration. (b)
Four rolls convection pattern
disappears when we increase the frequency to $\omega = 15.0$. Instead, the
time average of the one period of oscillation  is very random and the 
magnitude of the convection decays as the function of the periods, where
in the calculation, $\alpha = 0.5$, $\beta = 0.5$.

\vskip 1.0 true cm
\noindent {\bf III. Linear Stability Analysis of Traffic Model}
\vskip 0.5 true cm
Before studying the time dependent properties of the vibrating beds,
any model must make sure that the fixed bed solution, which is the
ground state at T=0 or equivalently $\Gamma =0$ [23], must be stable.
\vskip 0.5 true cm
(a) Fixed bed solution: Eqs.(1)-(3) assume two different fixed bed 
solutions.  The first one represents a uniform granular block
characterized by the homogeneous density, $\rho=\rho_c$ such
that $V(\rho_c)=0$, and ${\bf v} =0$.
But there is a second {\it non-uniform} solution, given by eq.(7).  Both 
solutions represent fixed bed solutions.  We only consider the stability of
a uniform solution.  Let
$\rho=\rho_c+\rho_1$ and ${\bf v} = {\bf v_1}$.  Substituting these into
eqs.(1)-(3), we find that $\rho_1$ satisfies the second order equation in time:

$$ \ddot \rho_1 + \dot\rho_1/\tau - \mu\nabla^2\dot\rho_1 - T_e\nabla^2\rho_1
=0 \eqno (8)$$

Assuming, $\rho_1=\rho_q(t) sin(\hat \pi mx) e^{iqz}$, we find:
$$\ddot \rho_q + B_o(q)\dot\rho_q + D_o(q)T_e\rho_q=0 \eqno (9)$$
with
$$B_o(q)=1/\tau + \mu(\hat\pi^2m^2 + q^2) \eqno (10a)$$
$$D_o(q)=T_e(\hat\pi^2m^2+q^2)\eqno (10b)$$

For $\rho_q(t) \sim e^{\sigma t}$, it is straightforward to
show that $Re\sigma$ is always negative, i.e,$Re\sigma <0$.  Thus,
the uniform fixed bed solution is stable.  
For the stability analysis of a nonuniform solution, see Appendix B.
\vskip 0.2 true cm

(b) Stability of the bouncing solution:  
When the system is subjected to vibrations, how do uniform and nonuniform
solutions
evolve to eventually produce convective rolls?  The standard way of
proceeding the stability analysis is first to find the basic
solution, and then exmaine its stability
by performing the linear stability analysis of this
basic solutions.  The onset of convection may be identified as a point
at which such basic solution becomes unstable.
In the thermal convection problem such
as the Rayleigh-Bernard convection, the basic solution is the laminar solution
with a temperature dependent density profile [25].  Since there exist two
fixed bed solutions, we need to study the stability of these two solutions
separately.
In what follows, we present only the stability analysis
associated with the uniform fixed bed solution,
and show in the Appedix B why the similar analysis for the nonuniform solution
is nontrivial.  Our analysis in what follows is based on the
belief that
since there is no difference between the two deep inside the
bed, the mechanism for the convection for the uniform and nonuniform 
solutions might be the same.
When $\Gamma\ne 0$, the basic solution associated with the uniform fixed
bed solution is also a {\it homgeneous} solution in density, i.e. $\rho=
\rho_o$, which
then undergoes harmonic oscillations.   This represents a uniform
granular block, represented by a single ball [25], moving up and down 
bouncing at the plate.  We previously
termed such a solution, a {\it bouncing solution} and such
a bouncing solution can be readily found by inspection
of the traffic equations (1)-(3).  Its horizontal velocity is zero, but the
vertical velocity, $\omega_o(t)$, is given by

$$v_z=\omega_o(t)= -V_o(\rho_o) + \hat V(\rho_o,\tau,\Gamma)(sin(\omega t)
-\omega\tau cos(\omega t)) \eqno (11)$$
where $\hat V(\rho_o,\tau,\Gamma)=\Gamma V_o(\rho_o)/(1+\omega^2\tau^2)$.
The numerically observed bouncing solutions
in Fig.2 may represent such a uniform bouncing solution.  

We now examine the bouncing solution in some detail, which is
essentially identical to a single ball dynamics, if we
focus on the motion of the center of mass of the block.  The dynamics of
a ball on a vibrating plate is non-trivial for high vibration strength, but
its trajectory obeys the simple Newtonian dynamics for low $\Gamma$ with
basically two characteristics: (i) the ball moves together with the plate
for some time, (ii) the ball is launched to the space at a particular
time $t_o$ that {\it depends} on the vibration strength.  We assume that
at the time of launching, the relative velocity between the block and the
plate is such that the relative distance, $\Delta (t)$,
between the ball and the plate starts
to increase at the time of launching.
Hence, $t_o$ is determined by the condition:
$$ \frac{\Gamma}{1+\omega^2\tau^2}(sin(\omega t_o)-\omega\tau cos(\omega t_o))
=1 + \epsilon \eqno (12)$$
where $\epsilon$ is determined by imposing the self-consistency
$\Delta (t=t_o+)>0$ (eq.14).
Note that (12) ensures that
$$ (1+\epsilon +\omega^2\tau^2)^{1/2} \le \Gamma \eqno (13)$$
For $\Gamma$ to be order one, $\tau$ needs to be of order $\sim 1/\omega$.
Next, since the velocity is given, we find that
the position of the ball in the moving frame, $\Delta (t) = \int_{t_o}^t
\omega(t')dt'$, is given by:
$$ \Delta (t) = -V_o(\rho_o)(t-t_o) + \frac{\Gamma V_o}{1+\omega^2\tau^2}
((cos(\omega t_o)-cos(\omega t))/\omega - \tau(sin(\omega t)-
sin(\omega t_o)) \eqno (14)$$

In Fig.10 are
shown typical $v_z(t)$ and $\Delta (t)$ as a function of time.  Note
that $\Delta$ is the relative distance between the ball and the plate and thus
cannot be negative: $\Delta(t)$ is positive for $t_o \le t\le t_1$
where $\omega_o(t_o)=\Delta(t_o)=0$ and $\Delta(t_1)=0$.  $T_f=t_1-t_o$,
which is a function of the vibration strength $\Gamma$, 
is the flying time of the ball.  For a set of parameters used in
simulations, i.e $\mu=\omega=\tau=1.0$,
and $V_o(\rho_o)=0.5$, we find $(t_o,t_1)$ changes from $(2.356,4.867)$ with
$T_f=2.51$ for
$\Gamma=2$ to $(1.197,5.0368)$ with $T_f=3.84$ for $\Gamma = 5$.  For
larger $\Gamma$, the flying time $T_f$ obviously increases.

\vspace{.3in}
\par
\centerline{\hbox{
\psfig{figure=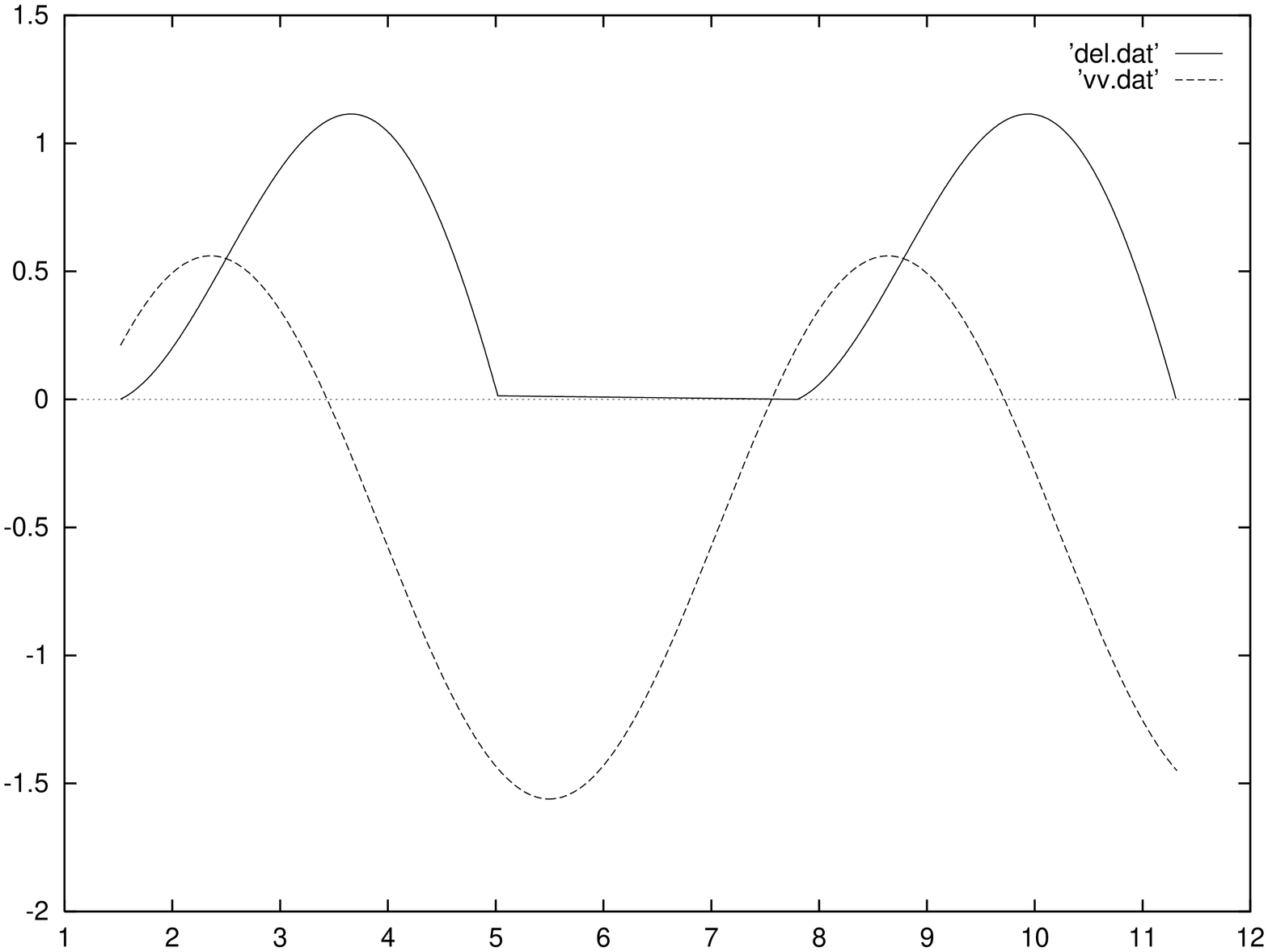,height=3.5in,width=3.5in,angle=270}
}}

Fig.10 Bouncing solutions.  The velocity,$\omega_o(t)$(lower one), 
and the position(upper one) of the ball
relative to the plate as a function of time.  The ball is launched at
$t=t_o$(Eq.14) with a positive velocity $v_z(t_o)>0$, and 
flies in the air until it
hits the plate again at $t=t_1$ with $\Delta(t_1)=0$.  
The ball stays at the plate and is
relaunched at $t=t_o+2\pi/\omega$.
\vskip 0.5 true cm
Let us now examine the stability of this bouncing solution.  Let
$\rho=\rho_o + \rho_L$ and {\bf v} = $\omega_o(t)\hat z + {\bf v_L}$ and
substitute $\rho$ and $v$ into equations (1)-(3).  Taking only the
linear term, we find:
$$ \partial_t\rho_L + \rho_o\nabla_{\xi}\bullet{\bf v_L}=0 \eqno (15a)$$
$$ \partial_t{\bf v_L} = \frac{T_e}{\rho_o}\nabla_{\xi}\rho_L
-{\frac{v_L}{\tau}} + \hat z{\frac{V_o'(\rho_o)}{\tau}}\rho_L
(-1+\Gamma sin(\omega t)) +\mu\nabla^2v_L \eqno (15b)$$
where $\xi = z-\int_o^t\omega_(t')dt'\equiv z 
-\Delta (t)$.  $\xi$ measures the point of the granular block
in the {\it box} frame that undergoes vibrations.  Now,
taking the divergence $\nabla_{\xi}$ of eq.(15b) and substituting $\rho_o
\nabla_{\xi}
\bullet v_L = -\partial_t\rho_L$, we find the equation for $\rho_L$:
$$ ({\it L}_o + {\it L}_1)\rho_L=0 \eqno (16)$$
where
$$ {\it L}_o= \partial^2_t - T_e\nabla_{\xi}^2 - \mu\nabla_{\xi}^2\partial_t
\eqno (17)$$
$${\it L}_1 = \tau^{-1}\partial_t + \rho_o\tau^{-1}V'_o(\rho_o)(
-1+\Gamma sin(\omega t))\partial_{\xi} \eqno (18)$$

In order to study the
stability of the bouncing solution, we now seek a solution of the form:
$$ \rho_L = \rho_{L,m,q}(t) sin(\pi mx/L) e^{iq(\xi + \Delta)}\eqno (19)$$
Note that m is integer and $\xi+\Delta \equiv z$.
Substituting (19) into (16), we obtain:
$$ \ddot\rho_q + B(q)\dot\rho_q + iC(q)\dot\rho_q + D(q)\rho_q + iE(q)\rho_q
=iL_q(t)\dot\rho_q(t) + M_q(t)\rho_q(t) + iN_q(t)\rho_q \eqno (20)$$
where
$$B(q) = B_o(q) = \mu(\hat\pi^2m^2+q^2)+1/\tau \eqno (21a)$$
$$C(q)=-2qV_o(\rho_o) \eqno (21b)$$
$$D(q) = D_o(q) -q^2(V_o^2(\rho_o) + \frac{\Gamma^2\hat V^2}{2}
(1+\omega^2\tau^2)\eqno (21c)$$
$$E(q)=q\mu(\hat\pi^2+q^2) - {\frac{\rho_o qV_o(\rho_o)}{\tau}}\eqno (21d)$$
$$ L_q(t)=-2q\Gamma\hat V (sin(\omega t) - \omega\tau cos(\omega t))
\eqno (21e)$$
$$ M_q(t) = q^2[{\frac{\omega^2\tau^2-1}{2}}\Gamma^2\hat V^2 cos(2\omega t)
-\omega\tau\Gamma^2\hat V^2 sin(2\omega t)]\eqno (21f)$$
$$N_q(t) = -\Gamma{\frac{\rho_o qV'_o}{2}}sin(\omega t) \eqno (21g)$$
where $\hat\pi=\pi/L$ with L the box size
and $\rho_q(t)=\rho_{L,m,q}(t)$.  For the most unstable mode, we set m=1.
We now change the time,
$t \rightarrow t_o+T$ and focus on the one oscillation from $T=0$ to
$T=2\pi/\omega$. Then, (21e)-(21g) are replaced by:
$$ L_q(t)\rightarrow  -2q\Gamma\hat V [(sin(\omega t_o)-\omega\tau 
cos(\omega t_o))cos(\omega T) + (cos(\omega t_o)+\omega\tau sin(\omega t_o))
sin(\omega T)] $$

$$ M_q(t)\rightarrow q^2\Gamma^2\hat V^2 [(\frac{\omega^2\tau^2-1}{2}
cos(2\omega t_o) - \omega\tau sin(2\omega t_o))cos(2\omega T) $$
$$ - (\frac{\omega^2\tau^2-1}{2} sin(2\omega t_o)+\omega\tau cos(2\omega t_o))
sin(2\omega T)] $$

$$ N_q(t) \rightarrow -\frac{\Gamma\rho_o qV'_o}{\tau} 
(sin(\omega t_o)cos(\omega T) + cos(\omega t_o) sin(\omega T)) $$

The stability of eq.(20), is highly nontrivial because it contains the
time dependent inhomogeneous term,
and we employ here an approximation method that was first used in ref.[8],
namely we  
replace the time dependent function $L_q(T),M_q(T)$ and
$N_q(T)$ by their {\it average} over one period, namely $<L_q(t)>=
T_f^{-1}\int_o^{T_f} dt L_q(T)$ with $T_f = (t_1-t_o)$, where 
the time dependent parts in
$L_q(t), M_q(t)$ and $N_q(t)$ in Eq.(21)
are replaced by their average values over the
{\it flying} time:
$$ sin(n\omega T) \rightarrow \frac{1}{n\omega T_f}(1-cos(n\omega T_f))$$
$$ cos(n\omega T) \rightarrow \frac{1}{n\omega T_f}sin(n\omega T_f)$$
where n=1 or 2.  The validity of such an approximation may be justified
since we are interested in the long time behavior and the system continuously
oscillates in time. 
Hence, we have replaced the {\it inhomogeneous} eq.(20) into the 
{\it homogeneous} one:
$$ \ddot \rho_q(T) + B(q) \dot\rho_q(T) + i\tilde C(q)\dot\rho_q(T)
+ \tilde D(q)\rho_q(T) + i\tilde E_q(T) \rho_q(T)=0 \eqno (22)$$
where 
$$\tilde C(q) = C(q)-<L_q(T)>\eqno (23a)$$
$$ \tilde D(q) = D(q)-<M_q(T)> \eqno (23b)$$
$$ \tilde E(q) = E(q) -<N_q(T)> \eqno (23c)$$

The stability of the homogeneous equation (Eq.22) with time independent
coefficients is easily determined by a standard linear stability analysis,
namely we set $\rho_q(t) = e^{\sigma t}$ 
and substitute this to (22) to obtain:
$$ \sigma^2+(B(q)+i \tilde C(q))\sigma+\tilde D(q)+ i \tilde E(q) =0 
\eqno (24)$$
The condition for the instability is when the perturbations grow in time,
i.e., $Re\sigma >0$.  Note that the solution of Eq.(24) is given by
$$\sigma_{\pm}
=-\frac{B(q)+i \tilde C(q)}{2}\pm 
\frac{1}{2}\sqrt{(B(q)+i \tilde C(q))^2-4 \tilde D(q)-4i \tilde E(q)}
\eqno (25)$$
In order to extract the real part, we notice that
the argument of the inside the square root may be written as $Re^{i\phi}$
with $R=((B^2-\tilde C^2-4\tilde D)^2) + 4B\tilde C -2\tilde E)^2)^{1/2}$
and $tan\phi = 2(B\tilde C-2\tilde E)/(B^2-\tilde C^2 -4\tilde D)$.
The real part of $\sigma_{\pm}$:

$$Re[\sigma_{\pm}]=-\frac{B}{2}\pm
\displaystyle\sqrt{\displaystyle\frac{1}{2}[\displaystyle\sqrt{
(\displaystyle\frac{B^2-\tilde C^2}{4}-\tilde D)^2+(\tilde E-\frac{1}{2}B 
\tilde C)^2}
+\displaystyle\frac{B^2-\tilde C^2}{4}-\tilde D]}\eqno (26)$$


When the real part
of eq.(26) becomes positive, the bouncing solution becomes unstable.
Hence, the instability condition for the most unstable branch becomes, 
$Re(\sigma_+)=-B/2+R^{1/2}cos(\phi/2)>0$, which is rewritten
by the following inequality:

$$ (B\tilde C-2\tilde E)^2\ge B^2(B^2-|B^2-\tilde C^2-4\tilde D|) \eqno (27)$$

Since the granular block is only in motion for the flying time, 
$0\le T\le T_f$, but sits on the plate for $T_f<T<2\pi/\omega$, the
stability of the granular block needs to be determined by the
{\it effective} instability condition, namely for the flying time, Eq.(27)
is operative, but for the rest of the time, the coefficients for the
fixed bed solutions, (10a) and (10b), must be used.  We thus define the
effective instability condition:

$$\sigma_{eff} = T_f[(B\tilde C-2\tilde E)^2 - B^2(B^2-|B^2-\tilde C^2
-4\tilde D|) $$
$$+ (2\pi/\omega - T_f)
(- B^2(B^2-|B^2-4D_o|) >0 \eqno (28)$$

When $\sigma_{eff} >0$, the bouncing solution becomes unstable.  
Since $B(0)=\mu\hat\pi^2 + 1/\tau$, $\tilde C(0)=C(0)-<L_q(0)>=0$,
$\tilde D(0)=D_o(0)-<M_q(0)>=D_o(0)=T_e\hat\pi^2$, 
$\tilde E(0) = E(0)-<N_q(0)>=0$,
we find 
$\sigma_{eff}(q=0)=-B(0)^2(B(0)^2-|B^2(0)-4D_o(0)|)\frac{2\pi}{\omega}$.
If $B^2(0)>4D_o(0)$ or $(\mu\hat\pi^2 + 1/\tau)^2>4T_e\hat\pi^2$,
then $\sigma_{eff}(0)<0$.  If $B^2(0)<4D_o(0)$, then $\sigma_{eff}(0)
=2B^2(0)(B^2(0)-2D_o(0))$ can be either positive or negative.  If
$\sigma_{eff}(0)>0$, then the bouncing solution is always unstable.  This
determines the criterion for the existence of the bouncing solutions in
the traffic equations:
$ (\mu\hat\pi^2 + 1/\tau)^2 < 4T_e\hat\pi^2$ and
$(\mu\hat\pi^2 + 1/\tau)^2 > 2T_e\hat\pi^2$.  And thus, if
$$(\mu+\hat\pi^2 + 1/\tau)^2 \le 2T_e\hat\pi^2$$
then, uniform bouncing solutions do not exist.  Otherwise, we expect the
bouncing solutions to appear. 

In Fig.11
is displayed $\sigma_{eff}$ as a function of the wave number q for 
$\tau=\omega$,$ \mu =1$, $\rho_o=0.53$, $V_o(\rho_o)=0.5$, 
$V_o'(\rho_o)=-10.$,
$L=32$, and $T_e=2.0$.  
Note that for $\Gamma <\Gamma_c$, $\sigma_{effc}$
is always negative, and thus the bouncing solutions are stable.  As we
increase the control parameter $\Gamma$, the peak of $\sigma_{eff}$ moves up
and becomes zero at the critical $\Gamma_c=2.27$, which signals the instability
of the bouncing solutions, and thus is identified as the onset of
convection.  
For $\Gamma>\Gamma_c$, 
there is a band of wavenumbers where
$\sigma_{eff}>0$.  The maximum growth rate is $q_c=1.11$ for
$\Gamma=2.32$ and $q_c=0.95$ at $\Gamma=2.27$, which are of order one.
The corresponding wavenumbers for the convective rolls are: 
$\lambda = 2\pi/k\sim$ 6.  Since there are two rolls in a box of
size L=32, the wavelength of the convective rolls is about 8, and our
analysis is consistent with the numerical results.  For other parameters,
the qualitative feature is the same, namely for $\Gamma<\Gamma_c$, 
$\sigma_{eff}<)$ for all q, and at a critical $\Gamma_c$, the local maximum
of $\sigma_{eff}$ crosses zero at $q=q_c$, and for $\Gamma>\Gamma_c$,
there is a band of wave numbers for which $\sigma_{eff}$ is positive.

\centerline{\hbox{
\psfig{figure=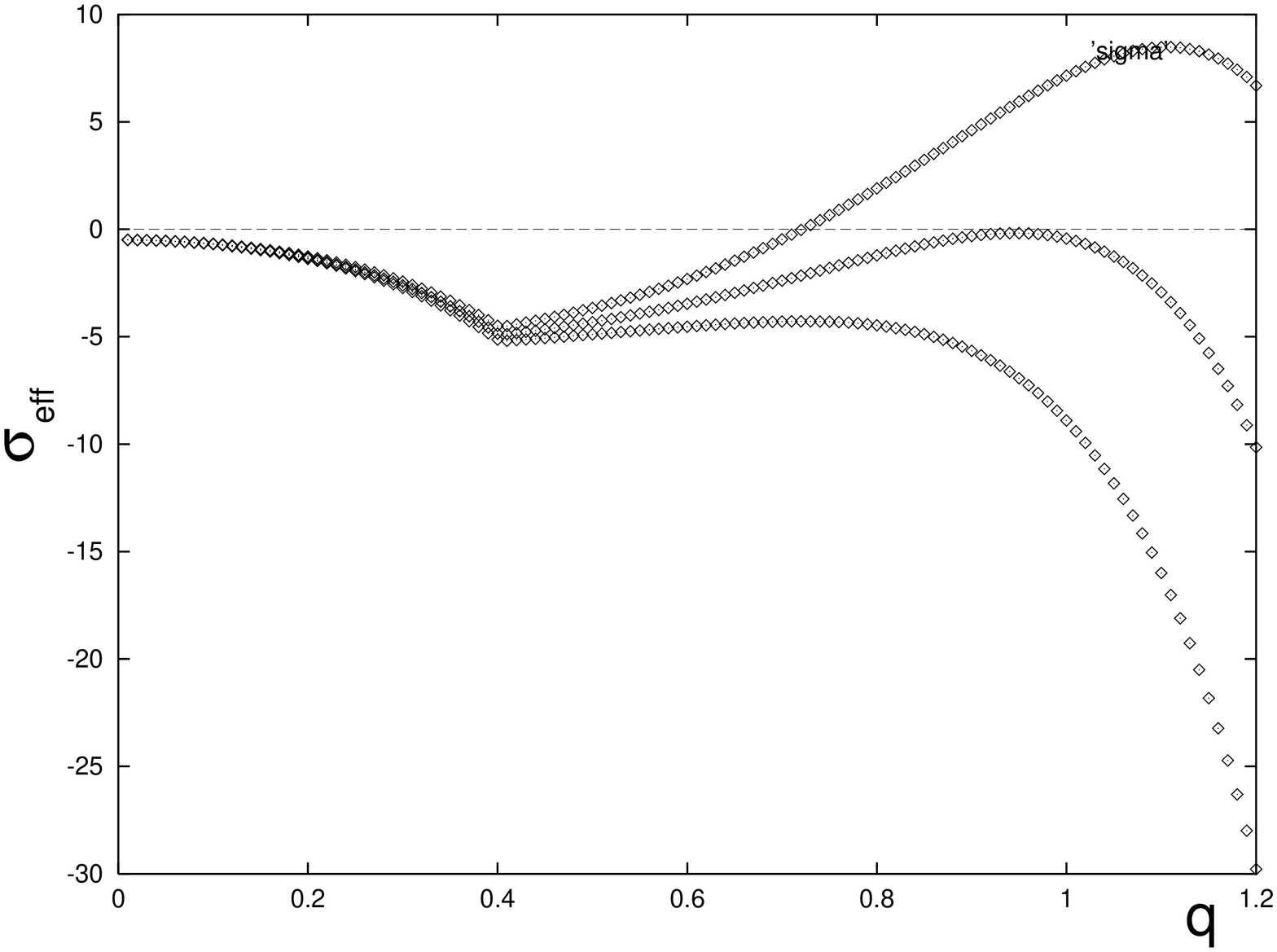,height=4.0in,width=4.0in,angle=270}
}}
\vspace{.3in}
Fig.11 The
effective instability condition $\sigma_{eff}(q)$(vertical axis)
is plotted against
the wave number q for different vibration strength $\Gamma$.
At the critical $\Gamma_c =2.27$, the local maximum of the growth rate crosses 
the zero.  For $\Gamma=2.2<\Gamma_c$, $\sigma_{eff}<0$ for all q, while
for $\Gamma_c<\Gamma=2.32$, there is a band of wave numbers for which 
$\sigma_{eff}>0$. 
\vskip 1.0 true cm
\noindent {\bf IV. Discussions}
\vskip 0.2 true cm
We conclude this paper by briefly reviewing the present status of the
convective instabiity of vibrating beds and simulations of the models
presented here (Model B) and elsewhere [1].
From our analytical as well as numerical analysis, we confirm that the 
bouncing motion of particle collections plays an important role in both model
A and model B.  We have, however, several issues which should be resolved in
the near future.

First, the role of boundary conditions.  We have employed no slip boundary 
conditions at the walls and the plate.  Further, we have put the rigid walls
at the top and thus suppressed the surface motion.  
Unlike the fluids, Granular materials
have been shown to exhibit very different motion near the wall in a zet flow 
experiment under gravity by Caram and Hong[20] 
and there has been some attempt to use the 
negative or positive slip to control the convective patterns [7].
More detailed studies to derive reasonable boundary conditions at the wall
are required.

Second, the role of interstitial fluid.  The model A[8] assumes no
interstitial fluid such as air, and it predicts a series of rolls for the 
vertically large aspect ratio.  Model B on the other hand predicts
that the convection suppresses in the bulk region but is
confined near the surface, which appears to be in accordance with
experiments and with the results of the two-fluid model with an interstitial
fluid [26].  Perhaps, the origin of drag, whether it is coming from
friction at the walls, or from the viscous effect of the interstitial fluid,
may not be relevant once it is present.  The suppression of the convection
in the bulk is due to the locking mechanism of grains
for near the closed packed density, which was taken into account in
the present model B by a cut off in $V(\rho)$, namely $V(\rho)=0$ for
$\rho>\rho_c$.  The results of model B, in particular the migration of
convective rolls toward the surface(Fig.4) and to the side walls(Fig.5)
seem to be more realistic and closer to the experimental and MD
simulation results than those of model A.
We certainly need more extensive studies of both model A and
model B to make quantitative comparison with experiments.

Third, future work must focus on completing the stability analysis
for the spatially nonuniform solutions(Appendix B) and its instability
mechanism along with the nonlinear analysis.
In addition, the study of the surface instability and its
connection to the surface fluidization and excitations 
as well as the study of horizontal vibrations and the onset of liquefaction
[9,27] remain important
open problems, which may shed insight into the oscillons [28] and other
related surface instabilities [29].  Such studies are 
currently under way and will be reported in future communications.
\vskip 2.0 true cm
\noindent {\bf Acknowledgment}
\vskip 0.2 true cm
DCH wishes to thank H. Hayakawa and D. A. Kurtze for helpful discussions over
the course of this work.

\newpage
\noindent {\bf Appendix A: Derivation of Traffic eq. from Enskoq eq.}
\vskip 0.2 true cm
We present here a brief derivation of the traffic equations from the
Enskoq equation.  
We first recall how one obtains the hydrodynamic eq. from Boltzmann eq.[21]
\vskip 0.2 true cm
$${\frac{\partial f}{\partial t}} + {\bf v}\bullet\nabla_rf + {\bf F}\bullet
\nabla_vf=J_B\eqno (A-1)$$
where F is the particle distribution function, and $J_B$ is the collision
integral.
If $\chi$ is summation invariant,i.e., then
$\chi_1+\chi_2=\chi'_1+\chi'_2$, where
' refers to the velocities of the two colliding particles after the
collision, and $\int d^3p\chi({\bf r},{\bf p})J_B=0$.
Now, one can derive hydrodynamic eqs. by multiplying 
$\int d^3p\chi({\bf r},{\bf p})$ to (A-1) as detailed in
Huang [30].  We now just focus on the Navier Stokes eq.
$$({\frac{\partial}{\partial t}} + {\bf u}\bullet\nabla){\bf u}
={\bf F}/m -{\frac{1}{\rho}}\nabla(P-{\frac{\mu}{3}}\nabla\bullet{\bf u})
+ {\frac{\mu}{\rho}}\nabla^2{\bf u}\eqno (A-2)$$

For granular materials, we need to take into account two
facts: first, particles are not point particles, but
hard sphere particles with finite diameter D, so one should use
Enskoq.eq. instead of Boltzmann eq [21].  Second,
we need to take into account gravity and friction,
for which the external force term {\bf F} in (A-2) may be replaced by
$$ {\bf F}/m = -g\hat z -\gamma{\bf u} \eqno (A-3)$$
The second term in (A-3) is due to the friction.  Further,
for Enskoq eq., there is collisional transfer.  So, the pressure tensor has
two components:
$$ P_{ij} = P_{ij}^{kinetic} + P_{ij}^{CT} \eqno (A-4)$$
The lowest order to the collisional transfer is: $P_{zz}=\chi(\rho)\rho^2/2$
with $\chi=(1-\rho/2)/(1-\rho)^3$, where $\chi$ is the two particle equilibrium
correlation function at the contact point of two colliding particles.  It
is known that the Enskoq pressure has the following two components:
$$ P = T\rho(1+\chi\rho/2)=P_{ideal} + P_1 \eqno (A-5)$$
Let us focus only on z component.  Then, the Navier-Stokes equation becomes:
$$({\frac{\partial}{\partial t}} + u_z{\frac{\partial}{\partial z}})u_z
= -g -\gamma u_z - {\frac{1}{\rho}}{\frac{d}{dz}}P+viscous \qquad
terms\eqno (A-6)$$
Now,
$$dP/dz = T[\rho'+\chi'\rho^2/2 + \rho\rho'\chi]$$
And the right hand side of (A-6) becomes:
$$R.H.S = -g -{\frac{T}{\rho}}{\frac{d\rho}{dz}} - T(\chi'\rho/2 + \rho'\chi) -
\gamma u_z$$
$$ = -{\frac{T}{\rho}}{\frac{d\rho}{dz}} + {\frac{(V(\rho)-u_z)}{\tau}}
\eqno(A-7)$$
where $\tau=\gamma^{-1}$ and
$$V(\rho) = -{\frac{T}{\gamma}}[\beta + \chi'\rho/2 + \rho'\chi]$$
\vskip 0.2 true cm
Hence, we have shown that
the z component of the Navier Stokes eq. reduces to the traffic eq.
$$({\frac{\partial}{\partial t}} + u_z{\frac{\partial}{\partial z}})u_z =
-{\frac{T}{\rho}}{\frac{d\rho}{dz}} + {\frac{(V(\rho)-u_z)}{\tau}} + 
\mu\nabla^2u_z \eqno (A-8)$$
For x and y components, simply use the Boltzmann eq.
\vskip 0.3 true cm
Note that within this derivation,
the mean speed has the following form:
$$V(\rho) = -{\frac{g}{\gamma}}{\frac{4-17\rho+8\rho^2}{4(1-\rho)^2}}$$
with the coefficient proportional to the gravity.  This is the reason for
the functional form taken in eq.(4) in the text.  For the equation of
void, we need to change $\rho\rightarrow (1-\rho)$, for which $V(\rho)$
decreases as a function of $\rho$.
\vskip 1.0 true cm
\noindent {\bf Appendix II: Stability analysis of a spatially
nonuniform fixed bed solution}
\vskip 0.5 true cm
We show here why the stability analysis of a spatially nonuniform
fixed bed solution is complicated.  In the absence of vibrations, $\Gamma=0$,
and the solution of this class, $\rho_o(z)$,
is given by eq.(7).  Since the solution is spatially inhomogeneous,
for small $\Gamma$, we seek for bouncing solutions that are only
a function of z:
$$ \rho(z) = \rho_o(z) + \rho'(z) s(z) e^{i\omega t} \eqno (B-1)$$
$$ v_z(z) = v(z) e^{i\omega t} \eqno (B-2)$$
where $\rho'(z) = d\rho(z)/dz$.  We put (B-1) and (B-2) into the traffic
equations (1)-(3) and linearize to obtain:
$$\frac{1}{\rho_o(z)}\frac{d\rho_o(z)}{dz}i\omega s(z) + 
(\frac{d}{dz} + \frac{1}{\rho_o}\frac{d\rho_o}{dz})v(z)=0 \eqno (B-3)$$
$$ - c_o^2\frac{\rho'_o}{\rho_o}\frac{ds(z)}{dz}-i\omega v(z)
-(1/\tau - \mu\frac{d^2}{dz^2})v(z) = \frac{\Gamma}{\tau}V(\rho_o(z))
\eqno (B-4)$$
Let $Q(z) = -\frac{1}{\rho_o}\frac{d\rho_o}{dz} = -\frac{1}{\tau c_o^2}
V(\rho_o)$.  Then, we find eqs. for Q(z) and v(z):
$$ -i\omega Q(z)s(z) + (\frac{d}{dz} - Q(z))v(z)=0 \eqno (B-5)$$
$$ -c_o^2Q(z)\frac{ds(z)}{dz} + (i\omega + 1/\tau -\mu \frac{d^2}{dz^2})v(z)
=c_o^2\Gamma Q(z) \eqno (B-6)$$
We need to find the solutions for (B-5) and (B-6) that both satisfy
the boundary conditions, $s(z), v(z)\rightarrow 0$ as $z\rightarrow \infty$,
which appears to be nontrivial if we relax the condition that $\Gamma$ is 
{\it not}
a small parameter. Even though we are lucky to find the solutions,
the stability analysis around these solutions still
present difficulties.
\newpage
\noindent {\bf References}
\vskip 0.2 true cm
\noindent [1] A short version of this paper was published in: H. Hayakawa
and D. C. Hong,' in Powders and Grains 97, 417 (1997), edited by
R. Behringer and J. Jenkins, Elsevier (New York). See also: cond-mat/970396.
and Su. Yue, ``'Nonlinear Phenomena in materials failure and granular
dynamics,' Ph. D thesis, Lehigh University (1995).

\noindent [2] M. Faraday, Phil. Trans. R. Soc. London {\bf 52}, 299 (1831).

\noindent [3] For recent review: see, H,M. Jaeger, S.R.Nagel, and R.P.
Behringer, rev. Mod. Phys. Vol 68, No.4, 1259 (1996)

\noindent [4] P. Evesque and J. Rajchenbach, Phys. Rev. Lett. {\bf 62},
44 (1989); C. Laroche, S. Douady, and S. Fauve, J. de Phys.(Paris)
{\bf 50}, 699 (1989); E. Clement, J. Duran and J. Rajchenbach,
{\it ibid}, {\bf 69}, 1189 (1992); J. Knight, H. Jaeger, and S. Nagle, 
{\it ibid} {\bf 70}, 3728 (1993);
H. Pak, E. van Doorn, and R. Behringer, {\it ibid}, {\bf 74}, 4643 (1995);
E.E. Ehrichs, H.M. Jaeger,
G.S. Karczmar, J.B. Knight, V.Y. Kuperman, and S. R. Nagel, Science {\bf 267},
 1632 (1995); H. Pak and R. Behringer, Nature, Vol 371, 231 (1994).

\noindent [5]J. Gallas, H. Herrmann and S. Sokolowski, Phys. Rev. Lett.
{\bf 69}, 1371 (1992); Y-h. Taguchi, {\it ibid} 1367.

\noindent [6] S. B. Savage, Adv. Appl. Mech. {\bf 24}, 289 (1984);
P. Haff, J. Fluid. Mech. {\bf 134}, 401 (1983).

\noindent [7] M. Bourzutschky and J. Miller, Phys. Rev. Lett. {\bf 74}, 221 
(1995).

\noindent [8] H. Hayakawa, S. Yue and D. C. Hong, Phys. Rev. Lett.
 {\bf 75}, 2328 (1995)

\noindent [9]T. Shinbrot, D. Khakhar, 
J. McCarthy and J.M. Ottino, Phys. Rev. Lett. {\bf 79}, 829 (1997);
K. Liffman, G. Metcalfe, and P. Cleary, {\it ibid} {\bf 79}, 4574 (1997)

\noindent [10] For example, see: C.S. Campell, Annu. Rev. Fluid. Mech.
{\bf 22}, 57 (1990).

\noindent [11] G. W. Baxter, R. Behringer, T. Fagert, and G. Johnson, Phys.
Rev. Lett. {\bf 62}, 2825 (1989).

\noindent [12] D. Helbing, {\it Verkehrsdynamik}(Springer, Berlin, 1997);
I. Prigogine and R. Herman, {\it Kinetic Theory of Behicular Traffic}
(Elsevier, New York, 1971).

\noindent [13] B. S. Kerner and P. Konhauser, Phys. Rev. E. {\bf 48}, 
R2335 (1993); {\bf 50}, 54 (1994), {\bf 51}, 6243 (1995); {\it ibid}
Phys. Rev. Lett. {\bf 79}, 4030 (1997).

\noindent [14] D. A. Kurtze and D. C. Hong, Phys. Rev. E. {\bf 52}, 218 
(1995).

\noindent [15] M. Bando et al, Phys. Rev. E. {\bf 51}, 1035 (1995).

\noindent [16] D. C. Hong, S. Yue, J. Rudra, M. Choi, Y. Kim, Phys. Rev. E.
{\bf 50}, 4123 (1994).

\noindent [17] 
O. Moriyama, N. Kuroiwa, M. Matsushita, and H. Hayakawa, cond-mat/9802174;

\noindent [18] H. Hayakawa and K. Nakanishi, cond-mat/9802131.

\noindent [19] T. Hill and J. Kakalios, Phys. Rev. E {\bf 49}, 3610 (1994)

\noindent [20]  H. Caram and D. C. Hong, Phys. Rev. Lett. {\bf 67}, 828 (1991);
Mod. Phys. Lett. B. 6, 761 (1992); For earlier development, see:
J. Litwinyszyn, Bull. Acad. Polon. Sci., Ser. Sci. Tech. {\bf 11}, 61 (1963);
W. W. Mullins, J. Appl. Phys. {\bf 43}, 665 (1972); J. App. Mech. 867 (1974).

\noindent [21] For example, see: S. Chapman and T.G. Cowling,
{\it The Mathematical Theory of Non-uniform Gases,} Cambridge Press (1958);
J. A. McLennan, {\it Introduction to Non-Equilibrium Statistical
Mechanics,} Prentice Hall, Englewood Cliff, NJ (1989).

\noindent [22] Y-h. Taguchi, Int. J. Mod. Phys. Vol. 7, Nos. 10
and 11, 1839 (1993)


\noindent [23] H. Hayakawa and D. C. Hong, Phys. Rev. Lett.
{\bf 78}, 2764 (1997)

\noindent [24] See for example: P. Manneville, {\it Dissipative Structures and
Weak Turbulence,} Academic Press (San Diego) (1990).

\noindent [25] For a single ball dynamics, see: J.M. Luck and A. Mehta, Phys.
Rev. E {\bf 48}, 3988 (1993) and references therein.

\noindent [26] H. Hayakawa and S. Sasa, Europhys. Lett. {\bf 17}, 685 (1992)

\noindent [27] S. Tennekoon and R. Behringer, patt-sol/9704001; D.E.
Rosenkranz and T. Poschel, cond-mat/9712108.

\noindent [28] F. Melo, P.B. Umbanhowar, and H.L. Swinney,
Phys. Rev. Lett. {\bf 75}, 3838 (1995);{\it ibid} Nature {\bf 382}, 793 (1996);
T.H. Metcalf, J.B. Knight, and H.M. Jaeger, Physica A {\bf 236}, 202 (1997);
K.M. Aoki and T. Akiyama, Phys. Rev. Lett. {\bf 77}, 4166 (1996).

\noindent [29] J. Eggers and H. Riecke, patt-sol/9801004; L. Tsimring and I. 
Aranson, Phys. Rev. Lett. {\bf 79}, 213 (1997)

\noindent [30] K. Huang, {\it Statistical Mechanics,}, p.96,
John Wiley and Sons (1987)
\end{document}